\documentclass[a4paper,fleqn]{cas-sc}
\usepackage[authoryear,longnamesfirst]{natbib}
\usepackage{graphicx}
\usepackage{listings} 
\usepackage[T1]{fontenc}
\usepackage{xcolor}
\usepackage{color}
\usepackage{longtable}
\usepackage[normalem]{ulem}

\usepackage{xcolor}
\usepackage{supertabular}
\usepackage{booktabs}
\usepackage{tikz}
\usepackage{enumitem}
\definecolor{dkgreen}{rgb}{0,0.6,0}
\definecolor{gray}{rgb}{0.5,0.5,0.5}
\definecolor{mauve}{rgb}{0.58,0,0.82}

\usepackage{caption}

\usepackage{listings}
\usepackage{xcolor}
\usepackage{hyperref}
\usepackage{url}

\usepackage{soul}
\usepackage{booktabs}
\usepackage{color}
\usepackage{tikz}
\usepackage{enumitem}

\lstdefinelanguage{Solidity}{
  keywords=[1]{anonymous, assembly, assert, balance, break, call, callcode, case, catch, class, constant, continue, constructor, contract, debugger, default, delegatecall, delete, do, else, emit, event, experimental, export, external, false, finally, for, function, gas, if, implements, import, in, indexed, instanceof, interface, internal, is, length, library, log0, log1, log2, log3, log4, memory, modifier, new, payable, pragma, private, protected, public, pure, push, require, return, returns, revert, selfdestruct, send, solidity, storage, struct, suicide, super, switch, then, this, throw, transfer, true, try, typeof, using, value, view, while, with, addmod, ecrecover, keccak256, mulmod, ripemd160, sha256, sha3}, 
  keywordstyle=[1]\color{blue}\bfseries,
  keywords=[2]{address, bool, byte, bytes, bytes1, bytes2, bytes3, bytes4, bytes5, bytes6, bytes7, bytes8, bytes9, bytes10, bytes11, bytes12, bytes13, bytes14, bytes15, bytes16, bytes17, bytes18, bytes19, bytes20, bytes21, bytes22, bytes23, bytes24, bytes25, bytes26, bytes27, bytes28, bytes29, bytes30, bytes31, bytes32, enu int, int8, int16, int24, int32, int40, int48, int56, int64, int72, int80, int88, int96, int104, int112, int120, int128, int136, int144, int152, int160, int168, int176, int184, int192, int200, int208, int216, int224, int232, int240, int248, int256, mapping, string, uint, uint8, uint16, uint24, uint32, uint40, uint48, uint56, uint64, uint72, uint80, uint88, uint96, uint104, uint112, uint120, uint128, uint136, uint144, uint152, uint160, uint168, uint176, uint184, uint192, uint200, uint208, uint216, uint224, uint232, uint240, uint248, uint256, var, void, ether, finney, szabo, wei, days, hours, minutes, seconds, weeks, years},  
  keywordstyle=[2]\color{teal}\bfseries,
  keywords=[3]{block, blockhash, coinbase, difficulty, gaslimit, number, timestamp, msg, data, gas, sender, sig, value, now, tx, gasprice, origin},  
  keywordstyle=[3]\color{violet}\bfseries,
  identifierstyle=\color{black},
  sensitive=false,
  comment=[l]{//},
  morecomment=[s]{/*}{*/},
  morecomment=[f][\lstbg{magenta!20}]{-\ },
  morecomment=[f][\lstbg{green!20}]{+\ },
  commentstyle=\color{gray}\ttfamily,
  stringstyle=\color{magenta}\ttfamily,
  morestring=[b]',
  morestring=[b]"
}

\makeatletter
\newenvironment{btHighlight}[1][]
{\begingroup\tikzset{bt@Highlight@par/.style={#1}}\begin{lrbox}{\@tempboxa}}
{\end{lrbox}\bt@HL@box[bt@Highlight@par]{\@tempboxa}\endgroup}

\newcommand\btHL[1][]{%
  \begin{btHighlight}[#1]\bgroup\aftergroup\bt@HL@endenv%
}
\def\bt@HL@endenv{%
  \end{btHighlight}%
  \egroup
}
\newcommand{\bt@HL@box}[2][]{%
  \tikz[#1]{%
    \pgfpathrectangle{\pgfpoint{1pt}{0pt}}{\pgfpoint{\wd #2}{\ht #2}}%
    \pgfusepath{use as bounding box}%
    \node[anchor=base west, fill=yellow!30,outer sep=0pt,inner xsep=1pt, inner ysep=0pt, rounded corners=0pt, minimum height=\ht\strutbox+1pt,#1]{\raisebox{1pt}{\strut}\strut\usebox{#2}};
  }%
}
\makeatother
\lstdefinestyle{Highlight}{
    moredelim=**[is][\btHL]{`}{`},
    moredelim=**[is][{\btHL[fill=orange!30]}]{´}{´},
    moredelim=**[is][{\btHL[fill=red!30]}]{@}{@},
}

\def\tsc#1{\csdef{#1}{\textsc{\lowercase{#1}}\xsp ace}}
\tsc{WGM}
\tsc{QE}

\begin{document}
\let\WriteBookmarks\relax
\def\floatpagepagefraction{1}
\def\textpagefraction{.001}

\title [mode = title]{An Empirical Study on Real Bug Fixes from Solidity Smart Contract Projects}

\author[1]{Yilin Wang}
\fnmark[1]
\ead{wangylin28@mail2.sysu.edu.cn}
 \ead[url]{https://orcid.org/0000-0002-0162-964X}
\affiliation[1]{organization={ School of Computer Science and Engineering,Sun Yat-sen University}, city={Guangzhou},country={China}}

\author[2]{Xiangping Chen}
\fnmark[2]
 \ead{chenxp8@mail.sysu.edu.cn}
  \ead[url]{https://orcid.org/0000-0001-8234-3186}
 \affiliation[2]{organization={School of Communication and Design,Sun Yat-sen University},city={Guangzhou},country={China}}

 \author[3]{Yuan Huang}
 \cortext[3]{Corresponding author}  
 \ead{huangyuan5@mail.sysu.edu.cn}
 \ead[url]{https://orcid.org/0000-0002-9548-0208}
 \affiliation[3]{organization={School of Software Engineering,Sun Yat-sen University},city={Zhuhai},country={China}}
\cormark[3]

 \author[4]{Hao-Nan Zhu}
 \fnmark[3]
 \ead{hnzhu@ucdavis.edu}
 \affiliation[4]{organization={Department of Computer Science, University of California, Davis}}
             
 \author[5]{Jing Bian}
  \fnmark[4]
   \ead[url]{https://orcid.org/0000-0002-3015-7573}
 \affiliation[5]{organization={School of Computer Science and Engineering,Sun Yat-sen University},
             city={Guangzhou},
             country={China}}
             
 \author[6]{Zibin Zheng}
 \fnmark[5]
  \ead{zhzibin@mail.sysu.edu.cn}
  \ead[url]{https://orcid.org/0000-0002-7878-4330}

 \affiliation[6]{organization={School of Software Engineering,Sun Yat-sen University}, city={Zhuhai},country={China}}
 
\begin{abstract}
Smart contracts are pieces of code that reside inside the blockchains and can be triggered to execute any transaction when specifically predefined conditions are satisfied. Being commonly used for commercial transactions in blockchain makes the security of smart contracts particularly important. Over the last few years, we have seen a great deal of academic and practical interest in detecting and fixing the bugs in smart contracts written by Solidity. But little is known about the real bug fixes in Solidity smart contract projects. To understand the bug fixes and enrich the knowledge of bug fixes in real-world projects, we conduct an empirical study on historical bug fixes from 46 real-world Solidity smart contract projects in this paper. We provide a multi-faceted discussion and mainly explore the following four questions: File Type and Amount, Fix Complexity, Bug distribution, and Fix Patches. We distill four findings during the process to explore these four questions. Finally, based on these findings, we provide actionable implications to improve the current approaches to fixing bugs in Solidity smart contracts from three aspects: Automatic repair techniques, Analysis tools, and Solidity developers.
\end{abstract}

\maketitle

\section{Introduction}

Blockchain uses cryptographic proof to replace trusted third parties to ensure the correctness of the information, allowing any two willing parties to transact directly with each other in an open environment. Ethereum \citep{wood2014ethereum} \citep{ethereum2014ethereum} is the most popular blockchain platform, which not only allows transactions with tokens but also offers storage and execution of the code, known as smart contracts. Smart contracts are at the core of Ethereum \citep{durieux2020empirical}, which are pieces of code that reside inside the decentralized blockchains in essence, and can be triggered to execute any task when specifically predefined conditions are satisfied \citep{gao2020deep}. Smart contracts in Ethereum are usually written using a statically-typed high-level programming language named Solidity \citep{Solidity}. As a young language born in 2015, Solidity is not safe enough and is exposed to severe bugs \citep{lutellier2020coconut}, which allow attackers to steal money or cause other damages while exploiting them. For example, in April 2018, the attackers exploited the Integer Overflow bug of the USChain BEC contract to copy tokens infinitely, causing the value of BEC tokens which is worth 900 million dollars to zero \mbox{\citep{BeautyChain}}. The attackers exploited the Reentrancy bug of SIREN AMM pools to steal nearly 3.5 million dollars worth of assets on 3 September 2021 \mbox{\citep{SIREN}}. Therefore, it is necessary to study the bugs in smart contracts in order to prevent such attacks and financial losses. \par

Over the last few years, we have seen a great deal of both academic and practical interest in the topic of detecting bugs in smart contracts written by Solidity developed for the Ethereum blockchain \citep{wang2013smartfixer} \citep{luu2016making} \citep{zou2019smart} \citep{perez2021smart} \citep{hwang2020gap} \citep{nguyen2021sguard} \citep{nguyen2013study} \citep{Mythril}. Meanwhile, \citep{nguyen2021sguard} \citep{chen2020defining} reduce the effort to repair the bugs in smart contracts by proposing various automatic repair techniques. Although quite a few works have focused on the bugs in Solidity smart contracts, the research community has limited knowledge of the naturalness of bug fixes in Solidity smart contract projects. For example, how many Solidity files are modified when fixing bugs? What is the most common fixed element in Solidity files during real bug fixes? It is of great importance to design an empirical study to answer these questions and to enrich the knowledge of real bug fixes in Solidity smart contract projects. The results may provide future insights for us to improve existing technologies.\par

In this paper, we focus on understanding bug fixes from real-world Solidity smart contract projects. We extract the bug fixes from the history of 46 Solidity smart contract projects and then conduct a multi-faceted empirical study to analyze them. In general, our research questions and their findings are as follows. \par
(1) \textbf{File Type and Amount.} We explore the types and the number of files involved during bug fixes to understand the distribution and complexity of bugs at the file level. Our results show that the distribution of bugs at the file level in Solidity smart contract projects is very complicated. Nearly 40\% of the bug fixes in Solidity files modify two or more Solidity source files. What is more, not only Solidity files but also the other source code files have quite a few bugs. About 92\% of the 6,146 bug fixes involve at least one the other source code file. 

(2) \textbf{Fix Complexity.} We explore the modifications of the Solidity code during bug fixes to understand the fix complexity in Solidity files. There are 19 element kinds that may be modified in Solidity files during bug fixes. Additionally, the fix actions taken to these elements contain \normalem{\emph{addition}}, \normalem{\emph{deletion}}, and \normalem{\emph{change}}. Specifically, we explore the modified elements, the element kinds, and the fix actions taken to Solidity files during bug fixes. Our results show that \normalem{\emph{Comment}} is the most common fixed element during bug fixes in Solidity files. What is more, the majority (59\%) of the Solidity files are multi-element modifications during bug fixes. \normalem{\emph{EventDefinition}} and \normalem{\emph{EmitStatement}} are the most relevant elements in the multi-element modifications with a correlation coefficient of 0.55. In general, a Solidity file involves an average of 2.5 code element kinds modification and 7.4 fix actions to code elements during bug fixes.\par
(3) \textbf{Bug Distribution.} We try to analyze the distribution of the bugs that can be detected by the current analysis tools Mythril and Slither in the historical bug-fixed versions of the Solidity files. Our results show that nearly 20.5\% of the Solidity files in our dataset have been exposed to bugs. They involve 14 categories of bugs. \normalem{\emph{Arithmetic}} and \normalem{\emph{Unused-return}} are the two most common bugs.\par
(4) \textbf{Fix Patches.} Based on the Bug Distribution, we want to find out how many bugs have been fixed, how many bugs have been newly introduced, and how developers fix them during the real bug fixes. The results show that the developers may not put much attention to fixing the bugs reported by the tools completely or avoid introducing them again. Meanwhile, Mythril and Slither perform poorly in detecting \normalem{\emph{Reentrancy}} bugs in practice. \par

\textbf{Contributions.} Our contributions in this paper can be summarized as follows. \par
(1) We provide a bug-fix dataset (commit-level) of 46 real-world Solidity smart contract projects. \par
(2) We provide a multi-faceted discussion of bug fixes in real-world Solidity smart contract projects.\par
(3) We analyze the bugs and their fixes in Solidity files from the history of 46 Solidity smart contract projects on GitHub.\par
(4) We provide actionable implications based on our results and findings for researchers to improve the automatic repair techniques and analysis tools.\par

The rest of this paper is structured as follows: Section 2 introduces the data extraction process and research questions we use to conduct the study. Section 3 presents the results of our empirical study and the actionable implications based on our findings. After that, Section 4 discusses the related work. Section 5 introduces the threats to validity. Finally, Section 6 gives the conclusions of this work. To facilitate research and application, our replication package and the dataset are available at \url{https://github.com/echowyl8/Real-Bug-Fixes-in-Smart-Contracts-Projects}.\par

\section{Methodology}
\subsection{Dataset}
To analyze the bug fixes from real-world Solidity smart contract projects, we first collect 50 projects from GitHub \citep{GitHub} in September 2021. In addition to downloading these 50 projects, we also download their commit metadata available. Then, we identify the bug fixes from these commit metadata. The specific data collection and screening are as follows.\par
\textbf{Search Key.} We collect the projects with the applied search key "contract" and "solidity" or "javascript". In this paper, we only explore the smart contract on Ethereum written by solidity. Because most of the smart contract platforms are JavaScript frameworks, we also consider the projects related to the keywords "contract" and "javascript". We select 25 projects in each of the two search results. All the 50 projects have been manually checked to assure they are Solidity smart contract projects. We invite three volunteer graduate students to conduct the validation process. They have an average of 1.5 years of development experience in Solidity smart contracts. Specifically, two participants independently determine whether the projects are related to Solidity smart contracts or not by reading the readme files of these projects and checking the Solidity files under their directories. If both of them agree that the project is related to Solidity smart contract projects in Ethereum, then the project is chosen. If there is a discrepancy between them, the rest participant determines the final result.

\textbf{Popularity.} The number of stars of a project is a proxy for its popularity on GitHub. Starring a project shows appreciation from the users\citep{wen2022quick}. We select the projects from the search results based on the number of stars.\par

\textbf{Bug Fix.} We follow previous work \mbox{\citep{lutellier2020coconut}} to extract bug fixes from the commit history of these 50 projects. We determine the commits as bug fixes if there are keywords "fix", "bug", and "patch" in their commit messages and filter commits using anti-patterns "rename", "clean up", "refactor", "merge", "misspelling", "compiler warning". According to the investigation of this approach in \mbox{\citep{lutellier2020coconut}}, it has a 93\% accuracy to extract bug fixes. \par

After filtering out projects with zero bug fix commits, we finally get 6,146 bug fixes from 46 projects for further study. These commits are created from February 2016 to September 2021. Figure 1 shows the number of non-bug-fix commits and bug-fix commits in the 50 projects. \normalem{\emph{contracts-solidity}} is the project with the most commits and bug-fix commits, which has 4,836 commits including 710 bug-fix commits. As shown in Table 1, there are 6,146 bug-fix commits in our dataset from these 46 projects. These projects involve 7,306 files with the total lines adding up to 2,686,281. Among them, there are 3,545 Solidity files with the total lines adding up to 241,445.\par

\begin{table}
\centering
\caption{Dataset statistics}
\begin{tabular}{@{}ccc@{}} 
\toprule Name & Amount \\
\midrule
The number of commits\hphantom{00} &\hphantom{0}43,354 & \\
The number of bug-fix commits\hphantom{00} &\hphantom{0}6,146& \\
The number of all the files\hphantom{00} &\hphantom{0}7,306& \\
The number of the Solidity files\hphantom{00} &\hphantom{0}3,545& \\
The lines of all the files\hphantom{00} &\hphantom{0}2,686,281& \\
The lines of the Solidity files\hphantom{00} &\hphantom{0} 241,445& \\
\bottomrule
\end{tabular}
\end{table}

\begin{figure}
	\centering
		\includegraphics[width=0.8\linewidth]{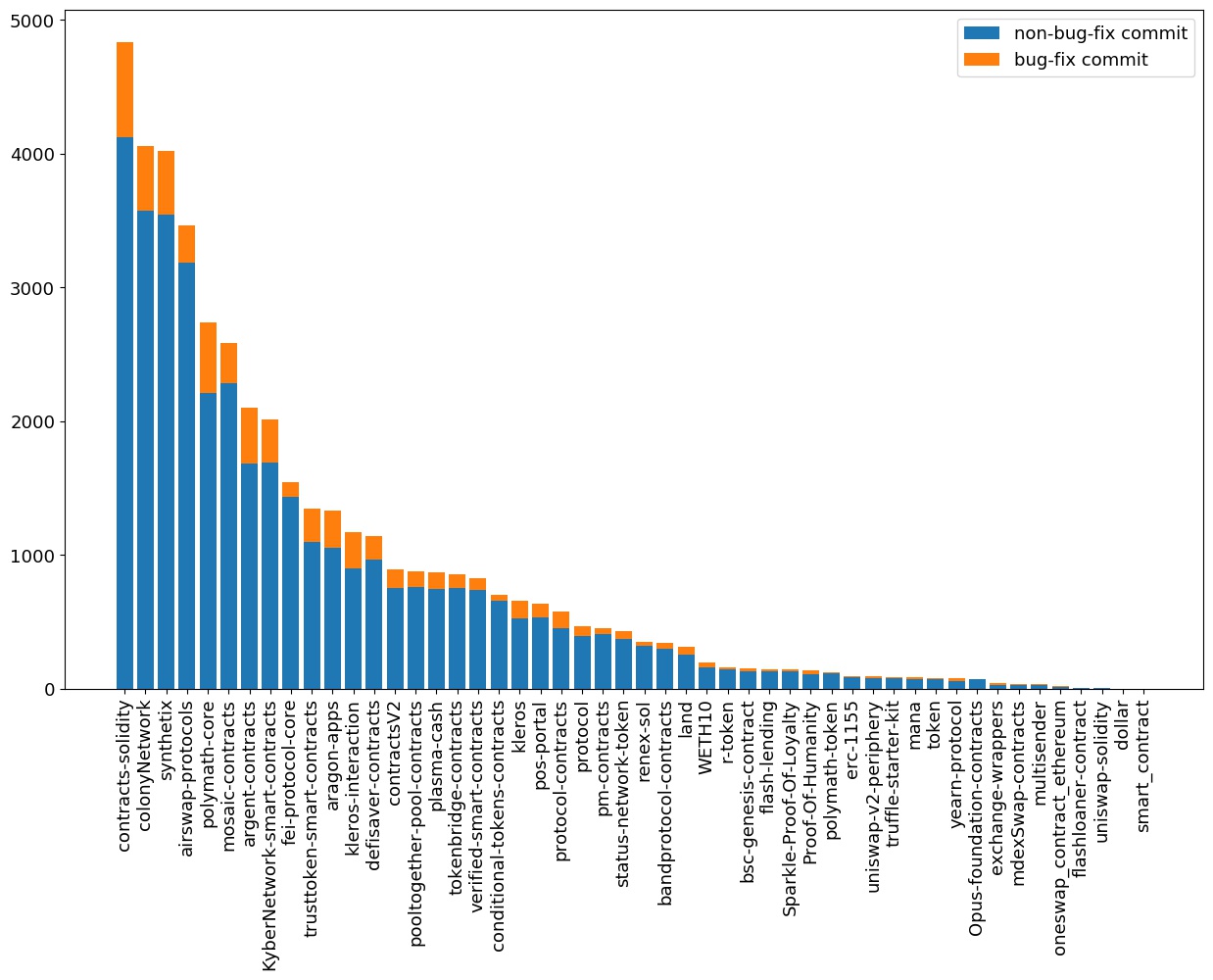}
	  \caption{The number of non-bug-fix commits and bug-fix commits of each project}\label{fig1}
\end{figure}

\subsection{Research Questions}
In order to understand the distribution of bugs and bug fixes in real-world Solidity smart contract projects. We consider the following four research questions. Their results can help developers understand the real bug fixes during the maintenance process of Solidity smart contract projects better.\par 

\textbf{File Type and Amount.} Recently, several automated smart contract repair approaches have been proposed \mbox{\citep{yu2020smart}} \mbox{\citep{nguyen2021sguard}} \mbox{\citep{chen2020defining}}. But they only modify one Solidity file so that the bugs which are not related to the Solidity files or involve more than one file could not be fixed by these automatic repair techniques. It is still unknown how many bugs in projects cannot be fixed by the limitations of these automatic repair techniques. So, in this part, we explore the types and the number of files involved during bug fixes to understand the distribution and complexity of bugs at the file level and to enrich the knowledge of bug fixes in Solidity smart contract projects. What is more, knowing the types and the number of files modified during bug fixes from real-world Solidity smart contract projects may help us to inspire novel approaches for finding, locating, and repairing bugs. 

The problems and goals in this part we study are as follows.\par
\textbf{RQ1.} What types of files are involved when fixing bugs? We try to find out the file types that are modified during bug fixes in real-world Solidity smart contract projects. It can enrich our knowledge of bug distribution at the file level in Solidity smart contract projects. \par
\textbf{RQ2.} How many Solidity files are modified during a fix? We try to find out the number of Solidity files modified during a bug fix. It can help us understand the impact of the bugs and the dependency among the Solidity files.\par
\textbf{RQ3.} How many Solidity files are necessary to be added or deleted to fix bugs? We try to find out the number of Solidity files added and deleted during a bug fix. It can help us to understand the need for adding and deleting Solidity files during a bug fix.\par

\textbf{Fix Complexity.} All the code elements such as \normalem{\emph{PragmaDirective}}, \normalem{\emph{FunctionDefinition}}, \normalem{\emph{RequireStatement}} and \normalem{\emph{VariableDeclaration}} in Solidity files can be modified during bug fixes. The non-code element \normalem{\emph{Comment}} also can be the modified object. Three types of fix actions can be taken to these elements during the fix: \normalem{\emph{addition}}, \normalem{\emph{deletion}}, and \normalem{\emph{change}}. For example, if the developers add a \normalem{\emph{RequireStatement}} during a bug fix. We consider that they take a fix action of the \normalem{\emph{ RequireStatement}} element. We want to explore the most common fixed element and the most common fix action taken to it. An Analysis of it can help us understand the bug fixes from real-world projects and find out the most susceptible element to bugs in practice. Further, we try to find out the element kinds and the number of fix actions that are involved during bug fixes to understand the complexity of fixes in Solidity files. The problems and goals in this part we study are as follows. \par

\textbf{RQ4.} What is the most common fix operation during bug fixes? In this question, we try to find out the most common fixed element and the most common fix action taken to it during the bug fixes. Such analysis can help us enrich the knowledge of the fixed objects and provide suggestions for future research of fixes.\par
\textbf{RQ5.} How many element kinds are modified in a Solidity file during bug fixes? We try to find out the number of element kinds that are involved in a Solidity file during bug fixes. It can help us understand the complexity of fixed objects during bug fixes at the code level in Solidity files.\par

\textbf{RQ6.} How many fix actions are taken to a Solidity file during bug fixes? We try to find out the number of fix actions taken to a Solidity file during bug fixes. It can directly show us the complexity of the fix in a Solidity file.\par

\textbf{Bug Distribution.}
Quite a few automated analysis tools have been proposed to detect the bugs in Solidity smart contracts. In this part, we try to use automated analysis tools to analyze the distribution of the bugs in the bug-fix versions of the smart contract projects. Such analysis can help us enrich the knowledge of bugs in real-world Solidity smart contract projects. The problems and goals in this part we study are as follows. \par

\textbf{RQ7.} What are the specific bugs reported by Mythril and Slither? \, We try to analyze the specific bugs that are reported by Mythril and Slither. It can help us understand the distribution of the specific categories of bugs in real-world Solidity smart contract projects. \par

\textbf{RQ8.} How many bugs are reported by Mythril and Slither and how many Solidity files do they exist in? We try to find out the number of bugs and the number of Solidity files that have been exposed to bugs from the historical bug-fix versions of the projects. 
\par

\textbf{Fix patches.}
Real bugs and their patches collected from real-world projects are critical for research in the repair of smart contracts. In this part, we want to find out the fixes of these bugs reported by Mythril and Slither, including the number of fixed bugs, non-fixed bugs, newly introduced bugs, and how they are fixed. Such analysis can help us understand the detailed fixed information of these bugs in real-world Solidity smart contract projects. The problems and goals in this part we study are as follows. \par

\textbf{RQ9.} How many bugs have been fixed and how many bugs have been newly introduced? We try to find out the number of the bugs that have been fixed and newly introduced during the bug fixes. It can help us to understand the detailed fixed information of these bugs in the maintenance of real-world projects.\par

\textbf{RQ10.} How do developers fix these bugs? 
We try to manually explore the human-written patches for these bugs.\par

\begin{figure}[]
	\centering
		\includegraphics[width=0.68\linewidth]{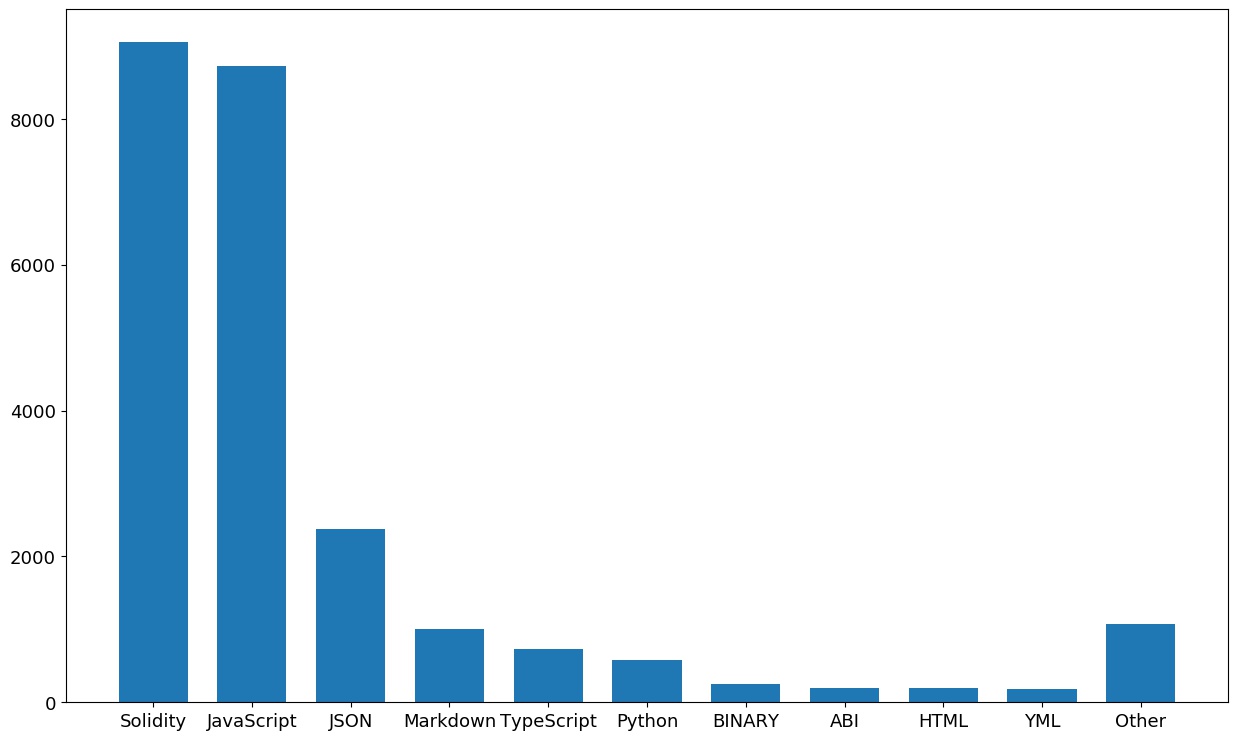}
	  \caption{The types of modified files during bug fixes}
\end{figure}

\section{Empirical Results}
In this section, we present the results of the empirical study and provide actionable implications based on our findings for researchers.\par
\subsection{File Type and Amount} 
\textbf{RQ1.} What types of files are involved when fixing bugs?\par 
We calculate the types of files that are involved during a bug fix. Figure 2 shows the top 10 types of modified files in the 6,146 bug fixes. As we can see from the results, except for Solidity files, JavaScript is the most common modified source code file type during bug fixes. TypeScript and Python are also among the common fixed code file types. What is more, there are almost as many bugs in JavaScript files as there are in Solidity files. The result is quite different from that in traditional projects like Java projects. Zhong et al. \mbox{\citep{zhong2015empirical}} conduct an empirical study on bug fixes from five popular Java projects. They find the most common modified files are Java files, and the other source code files are much fewer. It indicates that the bug fixes in smart contract projects are more complex than that in traditional projects. Because not only Solidity files but also the other source code files like JavaScript have quite a few bugs.\par
There are two general purposes for developers to use the other source code files in an Ethereum smart contract project. One is that the developers use programs written in various high-level languages such as JavaScript, TypeScript, and Python to develop clients to interact with Ethereum. In this way, to implement deploying and working with smart contracts, and to integrate with client nodes on the Ethereum network. Another is that the developers use these files for automated testing to validate the functionality of smart contracts from the perspective outside the blockchain. Both of the two usages require various operations to the code in the smart contract, which means that the bugs in these files may make these operations fail. For example in Listing 1, a test file \normalem{\emph{transferManager.js}} tries to test the function \normalem{\emph{cachedPrices}} in a Solidity file named \normalem{\emph{TokenPriceStorage}}. But \normalem{\emph{TokenPriceStorage}} does not have the function \normalem{\emph{cachedPrices}}, which fails the test. This bug fix is as follows, where the non-fixed code starts with \normalem{\emph{<}}, and the fixed starts with \normalem{\emph{>}}.\par

\begin{lstlisting}[caption={A fixed example in a JavaScript file}]
const TokenPriceStorage = require("../build/TokenPriceStorage");
let tokenPriceStorage = await deployer.deploy(TokenPriceStorage);
< const tokenPriceSet = await tokenPriceStorage.cachedPrices(erc20First.contractAddress);
---
> const tokenPriceSet = await tokenPriceStorage.getTokenPrice(erc20First.contractAddress);
\end{lstlisting}
The test file calls a function that does not exist and thus causes a bug. The fixed code changes the function call from \normalem{\emph{cachedPrices}} to \normalem{\emph{getTokenPrice}}, which is a function implemented in the Solidity file \normalem{\emph{TokenPriceStorage}}. As we can see from this example, the bug in the test file makes the test fail and thus the functionality of the smart contracts can not be tested. Similarly, the bugs in the deployment files may make the contract deployment fail and thus the project can not work properly. It indicates that we should pay attention to not only the bugs in Solidity files but also the other source code files that operate the Solidity files.\par
Among the non-source files, the most common modified files are JSON documents and Markdown files. The JSON files are used to exchange data and hold the project configuration. Solidity compilers use JSON files to capture the output for each compiled contract to store the contract metadata, including the link to the libraries, the deployed bytecode, the ABI, and so on. By using the information in these JSON files, developers can easily create objects in other languages to represent the smart contract and then use its functions. In our study, most of the modified JSON files are the latest outputs of the fixed smart contracts. While the Markdown files are manuals and tutorials which are used to explain something about the projects. The bugs in these non-source files may result in the usage of wrong data or operations by developers.\par
Further, we compare the numbers of Solidity files, the other source code files (except the Solidity files), and non-source files known as configuration files and natural language documents that are modified during a bug fix. Then, we calculate the percentage of bug fixes according to the number of these three kinds respectively. The results are shown in Figure 3. Its horizontal axes show the number of modified files in a bug fix and the vertical axes show the percentage of corresponding bug fixes. As shown in Figure 3, the number of bug fixes decreases with the increase in the number of modified files. In total, about 80\% of these bug fixes commits modify no more than one Solidity source file and nearly 20\% involve two or more Solidity files. Specifically, nearly 47\% (2,887) of the 6,146 bug fixes do not involve any Solidity files and 32\% of the 6,146 bug fixes involve one Solidity file. However, about 92\% of the 6,146 bug fixes involve at least one the other source code file, which is much higher than that of Solidity files. It indicates directly that bugs occur more frequently in the other source code files than in Solidity files. The developers mainly use the other source code files to interact with Ethereum and deploy the Solidity smart contracts, and test the functionality of the contracts. The bugs in these files may fail the deployment and testing of smart contracts, which may cause severe consequences. The high percentage of bug fixes that involve the other source files and the important functionality of these files inspire us to propose some novel tools in the future to help developers find the bugs in these files conveniently. In total, we can see that the distribution of bugs in smart contract projects is very complex, involving different source code files.\par

\begin{figure}[]
	\centering
		\includegraphics[width=0.68\linewidth]{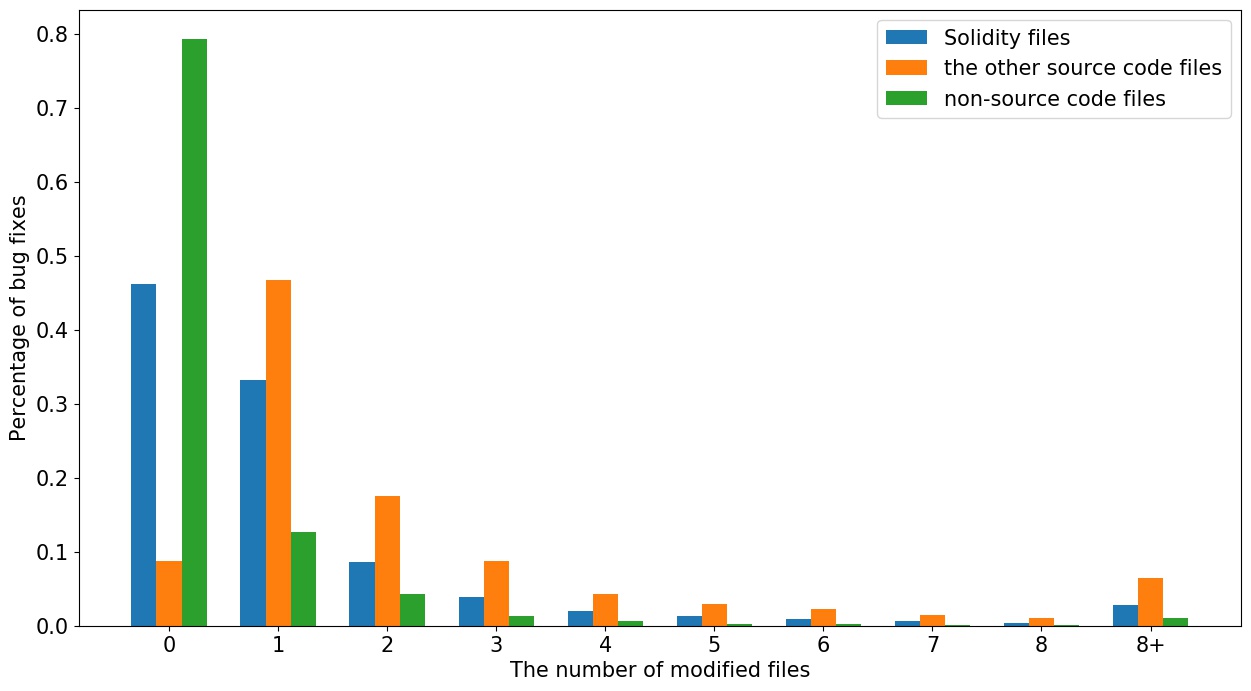}
        \caption{The number of files modified during a bug fix}
\end{figure}

\textbf{RQ2.} How many solidity files are modified during a fix? \par
In RQ1, we find that nearly 47\% (2,887) of the 6,146 bug fixes do not involve any Solidity files. In this RQ, we try to explore the number of modified Solidity files during the rest 3,259 bug fixes that involve Solidity files. To answer this question, we calculate the number of modified Solidity files during the 3,259 bug fixes. Figure 4 shows the results. Its horizontal axes show the number of modified Solidity files and the vertical axes show the percentage of the corresponding bug fixes. As shown in Figure 4, most bug fixes in Solidity files from these projects modify only one Solidity source file. But there are nearly 40\% of the 3,259 bug fixes that need to modify two or more Solidity source files. It shows the dependence of the Solidity files and indicates the complexity of the bug fixes at the Solidity file level. Current smart contract repair approaches \mbox{\citep{yu2020smart}} \mbox{\citep{nguyen2021sguard}} \mbox{\citep{chen2020defining}} can only repair one independent Solidity file, so these bugs could not be repaired by nowadays automatic program repair techniques. We can understand the bug fixes at the Solidity file level from this result and see the limitations of nowadays automatic repair techniques.\par

\begin{figure}[]
	\centering
		\includegraphics[width=0.68\linewidth]{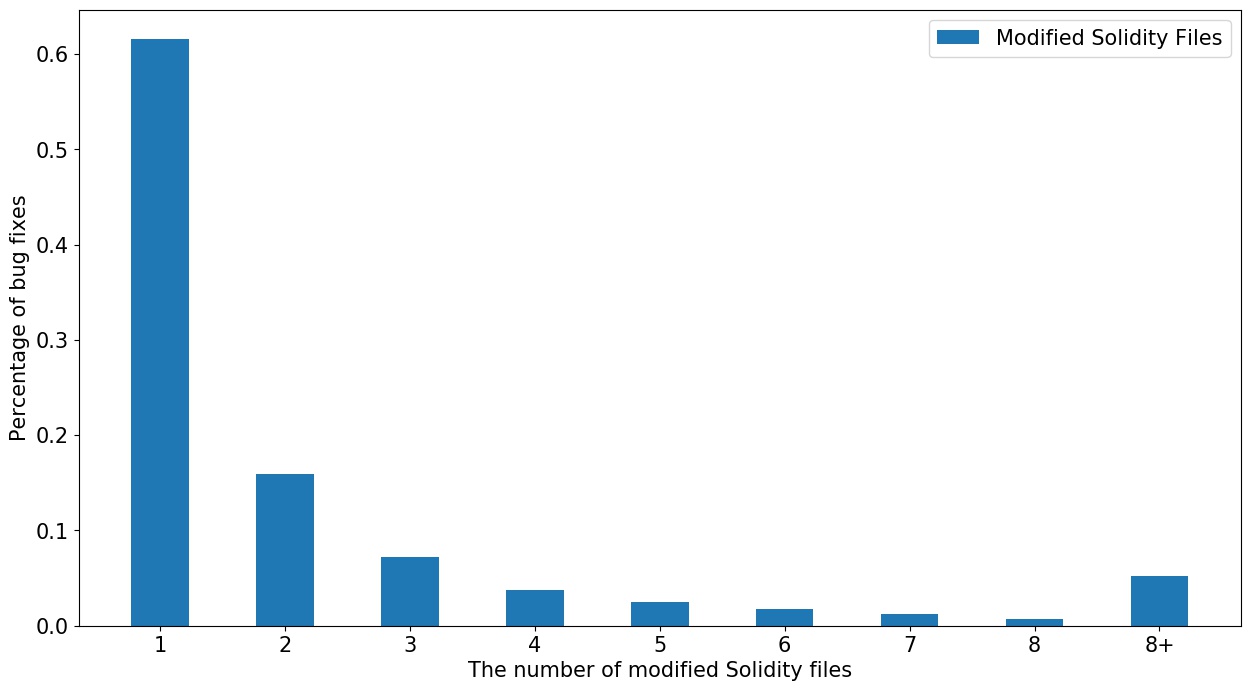}
	  \caption{The number of Solidity files modified during a bug fix}
\end{figure}

\textbf{RQ3.} How many solidity files are necessary to be added or deleted to fix bugs?\par
To answer this question, we calculate the number of Solidity files added and deleted during the 3,259 bug fixes that involve Solidity files. Then, we calculate the percentage of the bug fixes according to the number of Solidity files added and deleted and present the results in Figure 5. The horizontal axes show the number of Solidity files added and deleted, while the vertical axes show the percentage of the corresponding bug fixes. Generally, about 90\% of the 3,259 bug fixes do not need to add new Solidity files. Even if it needs to add files, one file (6\%) is quite enough. Almost all the 3,259 bug fixes that involve Solidity files do not need to delete Solidity files. Since most of the bug fixes that involve Solidity files do not need to add extra Solidity files or delete existing Solidity files, it is reasonable for us to focus on fixing codes in Solidity files.\par

The results lead to the following finding.\par
\textbf{\emph{Finding.1 }} The distribution of bugs in Solidity smart contract projects is very complex. About 92\% of the 6,146 bug fixes involve at least one the other source code file. Nearly 40\% of the 3,259 bug fixes that involve Solidity files modify two or more Solidity source files. 
\begin{figure}[]
	\centering
		\includegraphics[width=0.68\linewidth]{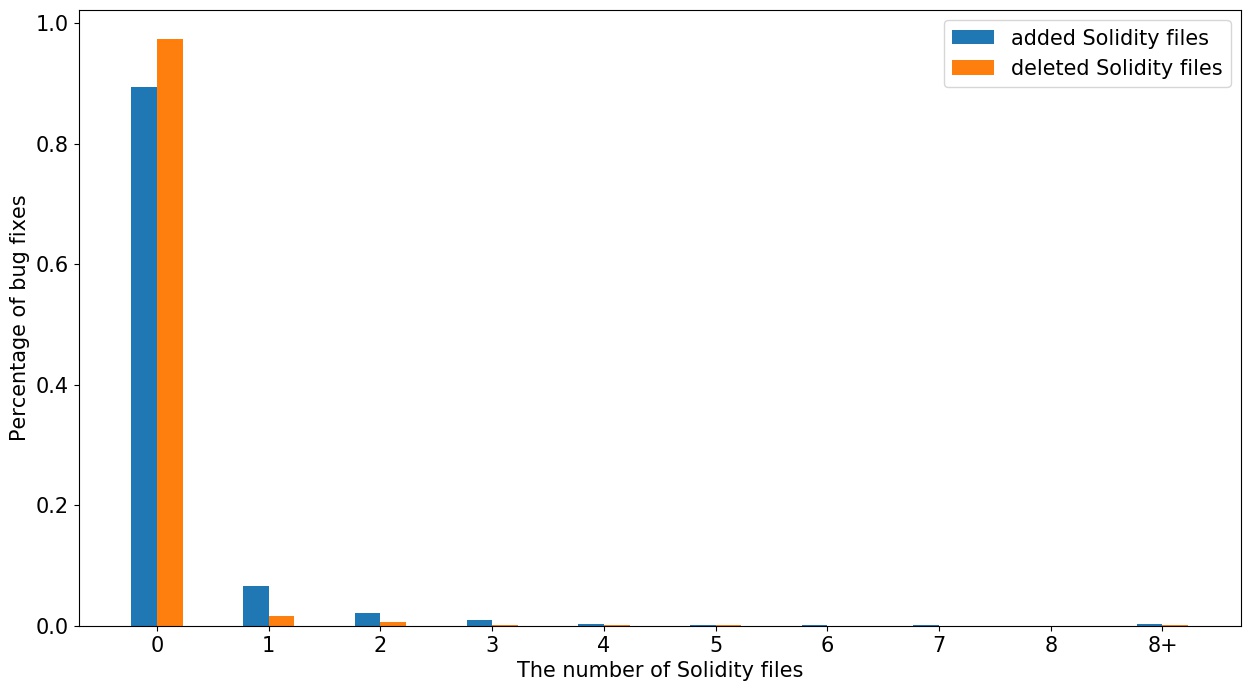}
	  \caption{The number of added and deleted Solidity files during a bug fix}
\end{figure}

\subsection{Fix Complexity}
After the discussion of file types and amount during bug fixes, in this section, we explore the complexity of these fixes at the Solidity code level.\par

\textbf{RQ4.} What is the most common fix operation during bug fixes?\par
In this paper, we use fix operation \normalem{\emph{<e, a>}} to describe the fix actions on different elements, where \normalem{\emph{e}} means an element and \normalem{\emph{a}} means one action of \normalem{\emph{addition}}, \normalem{\emph{deletion}}, and \normalem{\emph{change}}. Then, we calculate the number of fix operations in Solidity files to find out the most common one. We use AST to analyze and extract the fix actions to the code elements during bug fixes. The source code of each Solidity file is represented as an abstract syntax tree (AST) and we compare the ASTs of the origin Solidity file and the fixed Solidity file to get the fix operations during fixes. We do not consider the Solidity files that are added or deleted during bug fixes. For example, we count the number of \normalem\emph{FunctionDefinition} nodes before and after bug fixes to judge the addition and deletion of functions. Then, we consider the subnodes of the \normalem\emph{FunctionDefinition} nodes (the \normalem\emph{name}, the \normalem\emph{parameters}, the \normalem\emph{modifier}, the \normalem\emph{visibility}, the \normalem\emph{stateMutability}, the \normalem\emph{returnParameters}) to judge whether the \normalem\emph{FunctionDefinition} is changed or not. According to the grammar of Solidity \mbox{\citep{Solidity}} and the Solidity parser in Python \mbox{\citep{solidityparser}}, we finally define 19 kinds of code elements. As for the \normalem\emph{Comment}, we get their fix operations by comparing their number and content before and after the bug fix in Solidity files. Figure 6 shows the number of fix operations during bug fixes in Solidity files. Its vertical axis shows the names of the elements we define. Its horizontal axis shows the number of fix actions on \normalem\emph{addition}, \normalem\emph{deletion}, and \normalem\emph{change}.\par

As is shown in Figure 6, non-code element \normalem\emph{Comment} is the most common fixed element in our dataset. This result is not surprising, since it has the same result in traditional Java projects \mbox{\citep{zhong2015empirical}}. \normalem\emph{<Comment, addition>} is the most common fix operation during bug fixes in Solidity files. It is surprising that the number of comments that are added during bug fixes is so large. In Java projects \mbox{\citep{zhong2015empirical}}, the developers change the code comments much more frequently than adding them during bug fixes. But it is quite different in Solidity smart contract projects, the number of comments that are added during bug fixes is much larger than that are changed. It seems to indicate that developers like to add comments to the Solidity code during bug fixes. The maximum number of comments added in Solidity files appears in \normalem\emph{mosaic-contracts} with the commit hexsha \normalem\emph{4cca31ea4b2bc206f39a8553afba93a31e5c1bc7}, where the bug fix adds 435 lines of comments in one Solidity file. We manually check this modification and notice that most of the comments are added to explain the elements written before. The developers do not write comments to the code while they are implementing it but add them during bug fixes. There exists one research to help generate comments for Solidity files. Guang Yang et al. \mbox{\citep{yang2022ccgir}} propose an information retrieval-based code comment generation method for the functions of smart contracts written in Solidity. Since there are so many comments added to the code during bug fixes, it may still have space to improve the comment generation method in Solidity files.\par

\normalem{\emph{ExpressionStatement}} is the most common fixed code element. \normalem{\emph{<ExpressionStatement, change>}} is the most fix operation in code elements during bug fixes in Solidity files. The inside node \normalem{\emph{expression}} of \normalem{\emph{ExpressionStatement}} is complicated, including lots of types like \normalem{\emph{FunctionCall}}, \normalem{\emph{BinaryOperation}}, and so on. \normalem{\emph{FunctionCall}} may invoke complicated functions. \normalem{\emph{BinaryOperation}} may operate complicated objects. The complexity of the subnodes in \normalem{\emph{ExpressionStatement}} makes it much easier than the other elements like \normalem{\emph{EmitStatement}} that only has one type of subnode to be exposed to bugs and thus become the most common fixed code element. \normalem{\emph{FunctionCall}} is the most common fixed type in the subnodes of \normalem{\emph{ExpressionStatement}}. Maybe in the future, we can explore the invoked functions in \normalem{\emph{FunctionCall}} to learn more knowledge of the functions called during bug fixes in Solidity files. For example, we can analyze the bug fixes in the functions called from the open-source Solidity libraries. The result may give us suggestions when we fix the same type of bugs. Similarly, \normalem{\emph{FunctionDefinition}} also have complicated subnodes. \normalem{\emph{<FunctionDefinition, change>}} ranks the second among the fix operations in code elements. \normalem{\emph{FunctionDefinition}} has 6 types of subnodes that can be changed during bug fixes. We explore them to find out the most common fixed one. As is shown in Figure 7, \normalem{\emph{parameter}} is the most common fixed one in the changes of \normalem{\emph{FunctionDefinition}}. Specifically, nearly 51\% of the fix operations \normalem{\emph{<FunctionDefinition, change>}} modify the parameters of functions. For example in Listing 2, where non-fixed code statements start with \normalem{\emph{<}}, the fixed statements start with \normalem{\emph{>}}, and \normalem{\emph{c}} means change. To fix the bug, the name of the parameter \normalem{\emph{\_uri}} in the function \normalem{\emph{constructor}} is changed into \normalem{\emph{uri\_}}. All the usages of \normalem{\emph{\_uri}} also have to be modified. It is meaningful for future directions of IDE support to highlight the function calls while changing the functions and even automatically change the function calls to make it easier for developers to keep code consistent.

\begin{lstlisting}[language=Solidity, frame=none, breaklines=true, literate={\ \ }{{\ }}1,style=Highlight,caption={An example of modifying the parameter of a function}]
8c8
<     constructor(string memory _uri)
---
>     constructor(string memory uri_)
10,11c10,11
<         ERC1155(_uri)
<         NetworkAgnostic(_uri, ERC712_VERSION, ROOT_CHAIN_ID)
---
>         ERC1155(uri_)
>         NetworkAgnostic(uri_, ERC712_VERSION, ROOT_CHAIN_ID)
\end{lstlisting}


\begin{lstlisting}[language=Solidity, frame=none, breaklines=true, literate={\ \ }{{\ }}1,style=Highlight,caption={An example of \protect\normalem{\emph{Pragma Change}}.}]
3c3
<   pragma solidity >=0.6.2 <0.8.0;
---
>   pragma solidity 0.7.6;
\end{lstlisting}

\begin{figure}[]
	\centering
		\includegraphics[width=0.8\linewidth]{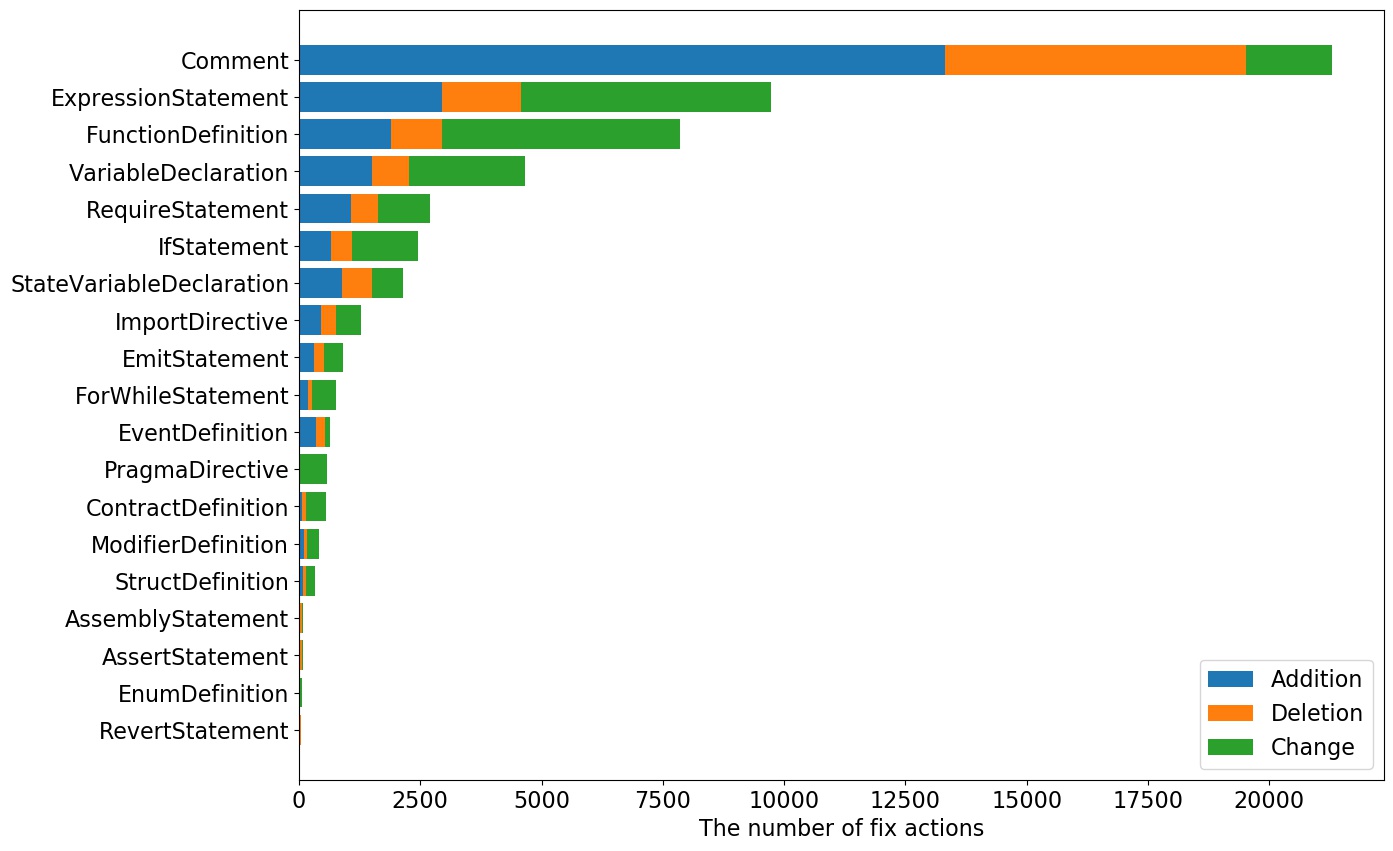}
	  \caption{Fix operations in Solidity files during bug fixes}
\end{figure}

\textbf{RQ5.} How many element kinds are modified in a Solidity file during bug fixes? \par
In RQ4, we have defined 19 kinds of elements that may be modified during bug fixes. In this RQ, we want to explore the number of element kinds that are involved in a Solidity file during bug fixes. First, we cluster Solidity files according to the number of element kinds that are modified during bug fixes and present the results in Figure 8. As is shown, during the bug fixes in Solidity files, 41\% of the Solidity files involve only one kind of element, which means that nearly 59\% of the Solidity files involve multi-element modification. More specifically, 20\% of the Solidity files involve two elements, and 12\% involve three-element. The number of Solidity files decreases with the increase in the number of element kinds. In general, a Solidity file involves 2.5 code element kinds on average during bug fixes. Further, we explore the kinds of elements that are modified. We cluster the Solidity files that only involve one kind of element and present the results in Figure 9. \normalem{\emph{Comment}} ranks the first. The result directly shows that quite a few bug fixes in Solidity files are taken only to fix the bugs in these code comments. It does not need to modify any code elements. Future research can explore it to help address the problems in these comments. \normalem{\emph{FunctionDefinition}} ranks the second among the single-element modification. Some modifications of the \normalem{\emph{FunctionDefinition}} may not cause the modifications of other code elements, such as the change of \normalem{\emph{visibility}} and \normalem{\emph{modifier}}. \normalem{\emph{PragmaDirective}} ranks the third. Quite a few Solidity files only involve the \normalem{\emph{PragmaDirective}} during bug fixes, which means the developers only change the versions of Solidity compiles. For example in Listing 3, to fix the bug in a Solidity file, the version of the pragma compile is changed from \normalem{\emph{ >=0.6.2 <0.8.0}} to \normalem{\emph{0.7.6}}. The version of the Solidity is changed from a range to an exact one. Most of the Solidity files only involving \normalem{\emph{PragmaDirective}} take the fix in the same way. It indicates that we have better use the exact version of compile to avoid the bugs caused by the versions of compiles.\par

\begin{figure}[]
	\centering
		\includegraphics[width=0.8\linewidth]{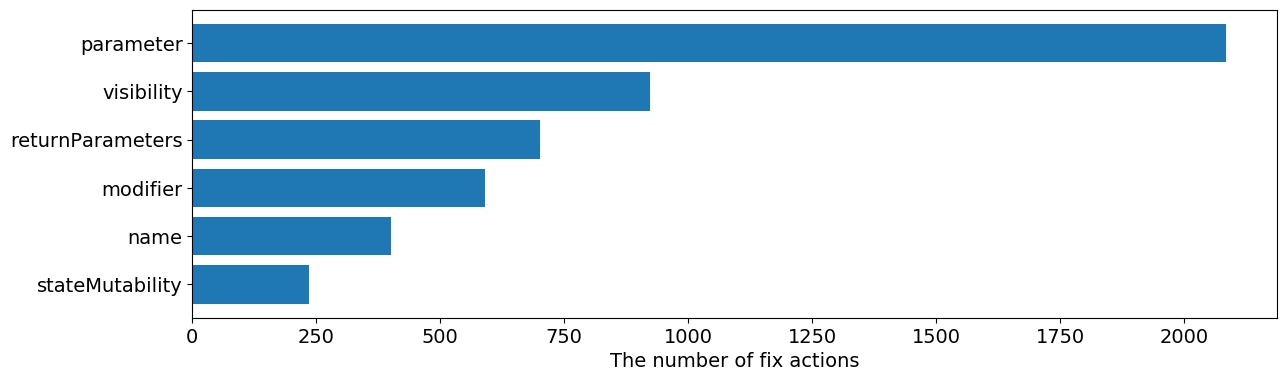}
	  \caption{Changes of \protect\normalem{\emph{FunctionDefinition}}}
\end{figure}

\begin{figure}[]
	\centering
		\includegraphics[width=0.8\linewidth]{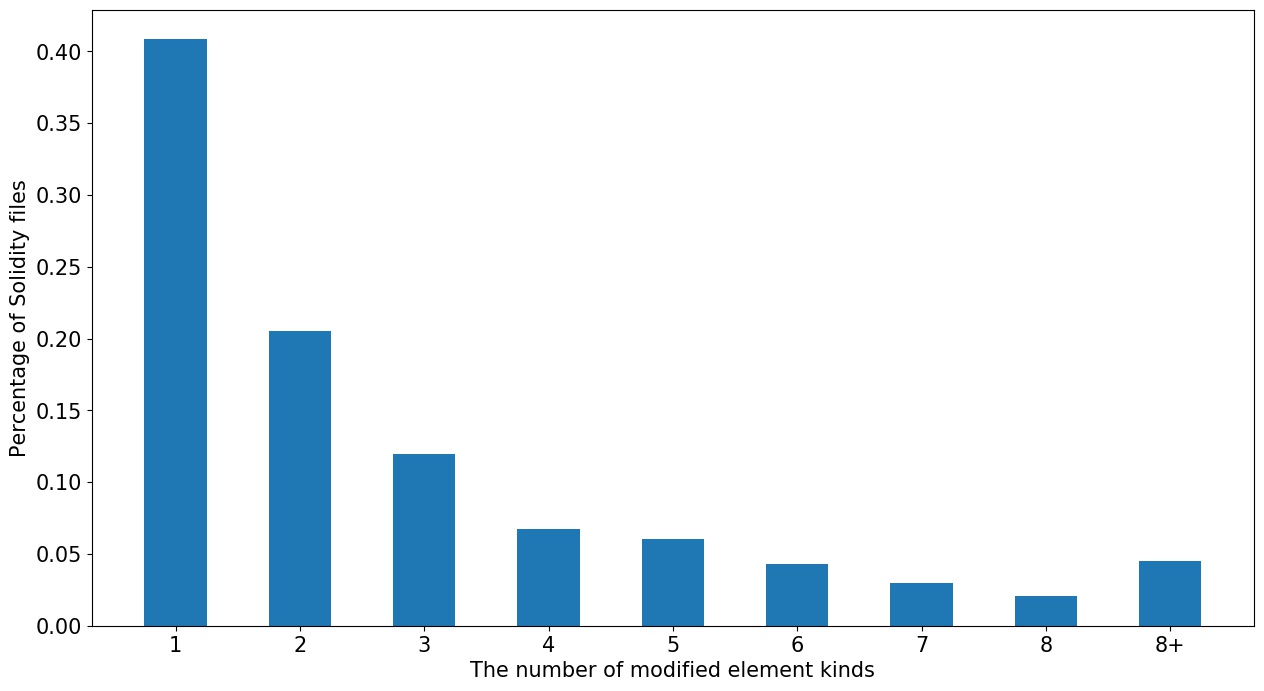}
	  \caption{The number of element kinds modified in a Solidity file during bug fixes}
\end{figure}
As for the multiple-element modification,  we use the Kindall \mbox{\citep{kendall1938new}} to analyze the correlation between elements. We show the results in Figure 10, where blue represents the positive correlation. The bluer it is, the stronger the positive correlation is. The results show that \normalem{\emph{EventDefinition}} and \normalem{\emph{EmitStatement}}  are the most relevant with a correlation coefficient of 0.55. \normalem{\emph{ImportDirective}}  and \normalem{\emph{ContractDefinition}}  rank the second with the 0.50. Event is used to store the arguments passed in transaction logs, which is called by \normalem{\emph{EmitStatement}}. So it is reasonable that the modification of \normalem{\emph{EventDefinition}} usually appears with the modification of \normalem{\emph{EmitStatement}}. It is meaningful for future IDE support to highlight \normalem{\emph{EventDefinition}} and its corresponding \normalem{\emph{EmitStatement}} to help the developers to manage the code. \normalem{\emph{ImportDirective}} is used to load other Solidity files for manipulation. Our result shows that the developers usually inherit the contracts in these imported Solidity files to fix bugs. The development between inherited and new contracts is also worth being researched.\par

\textbf{RQ6.} How many fix actions are taken to a Solidity file during bug fixes? \par
We calculate the number of fix actions that are taken to a Solidity file during bug fixes and cluster the Solidity files according to it. The result is shown in Figure 11. More specifically, we calculate the number of fix actions taken to all the elements and only to the code elements in a Solidity during bug fixes respectively. As is shown in Figure 11, nearly 31\% of the Solidity files are taken only one fix action to code elements. Combined with the result in RQ5, 41\% of the Solidity files modify one kind of element, we can conclude the result that most of the Solidity files that involve one element kind only need to be modified by one action. It is also what the existing repair techniques have done to fix. The result also shows that 68\% of the Solidity files need to be modified by several fix actions. The average number of fix actions to code elements and all the elements including \normalem{\emph{Comment}} in a Solidity file during bug fixes are 7.4 and 9.8. In RQ5, we notice that a Solidity file involves 2.5 code element kinds on average during bug fixes. It means that the developers need to take an average of three actions to one code element kind in a Solidity file during bug fixes. This result clearly shows the limitations of current automatic repair techniques, since the majority of fixes produced by them only take one action to one code element in a Solidity file. \par
\begin{figure}[]
	\centering
		\includegraphics[width=0.8\linewidth]{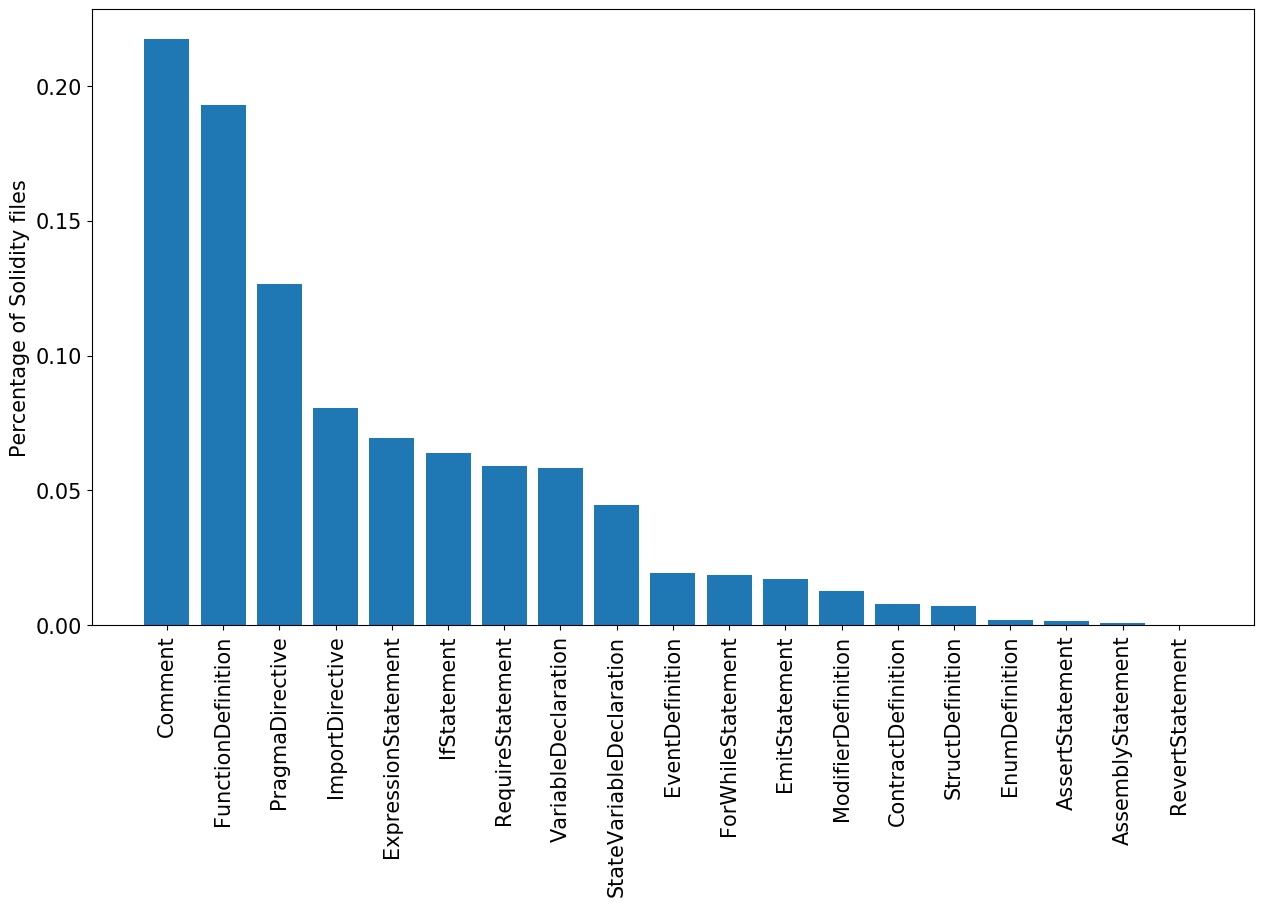}
	  \caption{The percentage of elements during one-element bug fixes}
\end{figure}

\begin{figure}[]
	\centering
		\includegraphics[width=0.68\linewidth]{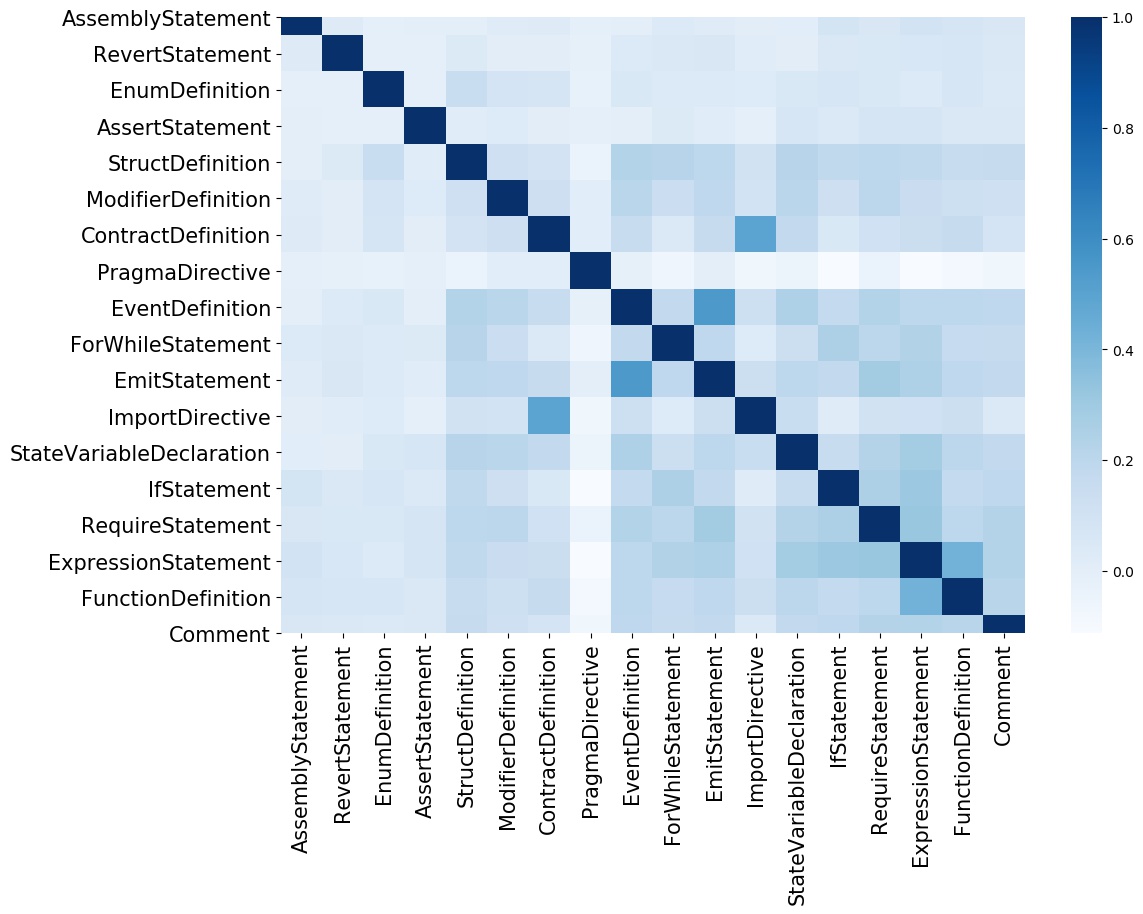}
	  \caption{The correlation between elements}
\end{figure}

\begin{figure}[]
	\centering
		\includegraphics[width=0.68\linewidth]{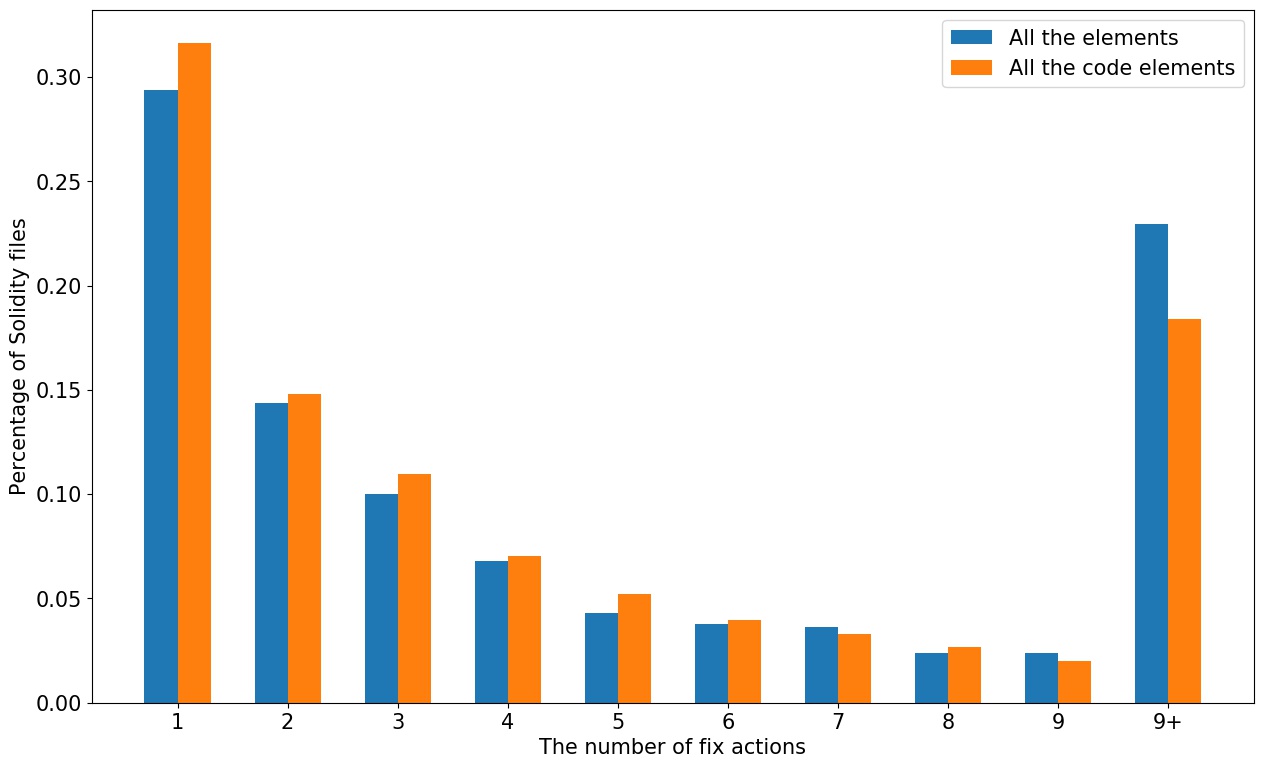}
	  \caption{The number of fix actions taken to a Solidity file during bug fixes}
\end{figure}

\textbf{\emph{Finding.2}} \normalem{\emph{Comment}} is the most common fixed element during bug fixes in Solidity files. Nearly 59\% of the Solidity files involve multi-element modification. \normalem{\emph{EventDefinition}} and \normalem{\emph{EmitStatement}} are the most relevant elements in the multi-element modification with a correlation coefficient of 0.55. On average, a Solidity file involves 2.5 code element kinds modification and 7.4 fix actions to code elements during bug fixes. \par


\subsection{Bug Distribution}
In this part, we try to analyze the distribution of the bugs that can be detected by analysis tools in the historical bug-fixed versions of the Solidity files. However, there is one problem we need to consider before our study. Due to the limitations of current analysis tools to detect bugs in Solidity files, we can not analyze Solidity files in these smart contract projects directly. The Solidity files in smart contract projects are much more complicated than those on Ethereum. These Solidity files may import Solidity files under other directories or online Solidity files to achieve certain functions. But current analysis tools to detect bugs in Solidity files can not identify and import the related Solidity files automatically. To solve this problem, we must import the dependent content of the Solidity files first. Then, use the analysis tools to detect the Solidity files with full imported content. In practice, we take three steps to solve this problem. First, we utilize the abstract syntax tree (AST) to obtain the imported files and the inheritance relationship among contracts from the Solidity files. Second, we use Topological Sorting \citep{kahn1962topological} to get the inheritance order of the contracts. Then, we extract the contracts from the Solidity files and merge them according to the inheritance order. Based on this method, we revert to the specific bug fix commits to obtain the Solidity files under different versions of the projects. We finally pack 116,410 Solidity files under the 3,259 bug-fix commits of the projects.\par

We use Mythril \mbox{\citep{mueller2018smashing}} and Slither \mbox{\citep{feist2019slither}} to detect bugs in Solidity files. According to the evaluation of nine current analysis tools \mbox{\citep{durieux2020empirical}}, the tool Mythril has the best accuracy among them. But Mythril is not powerful enough to replace all the tools. So we follow the suggestion in \mbox{\citep{durieux2020empirical}} and use the combination of Mythril and Slither.



\begin{longtable} {p{120pt}p{290pt}}
\caption{The description of the 14 categories of bugs}\label{tab:first} \\
\toprule 
category  &  Description\\ 
\midrule
 Arithmetic &  \normalem{\emph{Integer Overflow}} \citep{IntegerOv}, and \normalem{\emph{Integer Underflow}} \citep{IntegerUn}. \\
\midrule
 Reentrancy  & Reentrant function calls make a contract behave in an unexpected way \citep{durieux2020empirical}.
\\
\midrule

Front-Running   & Two dependent transactions that invoke the same contract are included in one block \citep{durieux2020empirical}.\\
\midrule
 Access Control   &  The developers do not restrict or incorrectly restrict the access of functions, use of tx.origin, or make reckless use of delegatecall \citep{durieux2020empirical}.\\
\midrule
 Bad Randomness  & Malicious miner biases the outcome \citep{durieux2020empirical}. \\
\midrule

Time manipulation   & The timestamp of the block is manipulated by the miner \citep{durieux2020empirical}. \\
\midrule

 Denial of Service   & The contract is overwhelmed with time-consuming computations \citep{durieux2020empirical}. \\
\midrule
 Locked-ether    & Ethereum smart contracts can receive Ether. But the received funds might get locked permanently into the contract \citep{perez2021smart}.\\
\midrule
Uninitialized variables   & Uninitialized variables.\\
\midrule
Strict Balance Equality  & Use of strict equalities that can be easily manipulated by an attacker \citep{feist2019slither}.\\
\midrule
Shadow variables    & 
Shadow variables occur within a single contract when there are multiple definitions on the contract and function level \citep{Shadowing}.\\
\midrule
Unmatched ERC-20 Standard    & The function name, parameter types, and return value do not strictly follow the ERC20 standard, for example, miss return values or miss some functions \citep{chen2020defining}. \\
\midrule
Unused-return & The return value of an external call is not stored in a local or state variable \citep{feist2019slither}. \\
\midrule 
\textit Arbitrary-send & Unprotected call to a function sending Ether to an arbitrary address, or when \normalem{\emph{msg.sender}} is not used as from in \normalem{\emph{transferFrom}}.
\\

\bottomrule
\end{longtable}

\textbf{RQ7.} What are the specific bugs reported by Mythril and Slither? \par
Considering the impact and severity of the bugs, we do not take into account the informational bugs reported by Mythril or Slither. The two tools detect 27 types of bugs in our dataset in total. Mythril and Slither use different descriptions for bugs. To facilitate our analysis, we first need to have a unified description of these bugs. We classify these bugs according to the taxonomy presented in the DASP10 \mbox{\citep{DASP}}. As for the bugs that DASP10 does not cover, we classify them according to their characteristic. We finally divide these 27 types of bugs into 14 categories. Table 2 presents detailed information about the categories.

\textbf{RQ8.} How many bugs are reported by Mythril and Slither and how many Solidity files do they exist in? \par
We identify the bugs by \normalem{\emph{<category, project, function, code>}}. Then, we calculate the number of bugs reported by the two tools. There are some categories like \normalem{\emph{Reentrancy}} and \normalem{\emph{Access Control}} that can be detected by both Mythril and Slither. We consider the bugs that are both reported by Mythril and Slither to reduce false positives. As for those categories that can only be detected by one tool like \normalem{\emph{Arithmetic}} and \normalem{\emph{Time manipulation}}, we directly calculate the number of bugs detected by the tool. Table 3 shows the number of bugs identified by Mythril and Slither. Each row in Table 3 represents a bug category. The Total column shows the number of bugs.\par
As Table 3 is shown, \normalem{\emph{Arithmetic}} is the most common bug category, which refers to the bugs \normalem{\emph{Integer Overflow}} and \normalem{\emph{Integer Underflow}}. \normalem{\emph{Integer Overflow}} and \normalem{\emph{Integer Underflow}} are common types of bug in many programming languages \mbox{\citep{perez2021smart}}. But they can be used by hackers to attack and steal money from smart contracts in the context of Ethereum. The \normalem{\emph{Unused-return}} is the second most common category in our dataset. It means that developers lack a check on the return value of a function. Without the check for the return value, developers can not detect unexpected states and conditions of these functions \mbox{\citep{unusedreturn}}. The code in the functions may resume executing even if the invoked function throws an exception. An attacker could force the function to fail or otherwise return a value that is not expected, then the subsequent program logic could lead to unexpected results.  \par
There are 3,545 Solidity files with different names from the 3,259 bug-fix versions of the 46 smart contract projects in total. After removing the duplicates, we find that nearly 20.3\% (720) of the 3,545 Solidity files have been exposed to bugs.

\textbf{\emph{Finding.3}} Nearly 20.3\% of the 3,545 Solidity files in our dataset have been exposed to bugs. There are 14 categories of bugs reported by Mythril and Slither. \normalem{\emph{Arithmetic}} and \normalem{\emph{Unused-return}} are the two most common bugs.\par

\begin{table}
\centering
\caption{Bugs identified per category by Mythril and Slither}
\begin{tabular}{@{}cccc@{}} 
\toprule category & Mythril & Slither  & Total \\
\midrule
 Arithmetic \hphantom{00} &\hphantom{0}803 & \hphantom{0}/ & \hphantom{0}803 \\

Reentrancy \hphantom{00} &\hphantom{0}859 & \hphantom{0}942 & \hphantom{0}141 \\

  Front-Running \hphantom{00} &\hphantom{0}52 & \hphantom{0}0 & \hphantom{0}52 \\
Access Control \hphantom{00} &\hphantom{0}20& \hphantom{0}17 & \hphantom{0}8 \\
Bad Randomness \hphantom{00} &\hphantom{0}27& \hphantom{0}0 & \hphantom{0}27 \\
Time manipulation \hphantom{00} &\hphantom{0}/& \hphantom{0}58 & \hphantom{0}58 \\
Denial of Service  \hphantom{00} &\hphantom{0}/& \hphantom{0}418 & \hphantom{0}418\\
Locked-ether \hphantom{00} &\hphantom{0}/ & \hphantom{0}51 & \hphantom{0}51 \\
Uninitialized variables \hphantom{00} &\hphantom{0}/& \hphantom{0}364 & \hphantom{0}364 \\
Strict Balance Equality \hphantom{00} &\hphantom{0}/& \hphantom{0}94 & \hphantom{0}94 \\
Shadow variables  \hphantom{00} &\hphantom{0}/& \hphantom{0}178 & \hphantom{0}178 \\
Unmatched ERC-20 Standard \hphantom{00} &\hphantom{0}/& \hphantom{0}30 & \hphantom{0}30 \\
Unused-return \hphantom{00} &\hphantom{0}/& \hphantom{0}470 & \hphantom{0}470 \\
 Arbitrary-send \hphantom{00} &\hphantom{0}/& \hphantom{0}119 & \hphantom{0}119 \\
\bottomrule
Total \hphantom{00} &\hphantom{0}1,761& \hphantom{0}2,741 & \hphantom{0}2,813 \\
\bottomrule
\end{tabular}
\end{table}
\subsection{Fix patches}
As we introduced in Bug Distribution, we check out 3,259 bug fix commits of the 46 projects and finally pack 116,410 Solidity files under these versions. In this part, we use these historical versions of Solidity files to determine whether the bugs that are reported by Mythril and Slither have been fixed or not, how many bugs have been newly introduced, and how they are fixed.\par
\textbf{RQ9.} How many bugs have been fixed and how many bugs have been newly introduced? \par
We use \normalem{\emph{<category, project, function, code>}} as the unique identifier IDs for bugs to facilitate our tracking of these bugs. It is worth noting that the code specifically refers to those that are capable of detecting bugs. During the process of fixing bugs, the developers may modify or add new code, then these changes may cause the identifier IDs of the bugs to change. For example, the developers change the code statement of the unique identifier ID, which causes the identifier ID to change from \emph{<category, project, function, code>} to \emph{<category, project, function, code\_c>}. In this case, we consider the original bugs \emph{<category, project, function, code>} to be fixed, but the developers may have introduced a new bug by \emph{code\_c}. Hence, simply monitoring the number of fixed and non-fixed bugs does not accurately reflect the level of significance that developers place on these bugs. To address this, we calculate the number of bugs that have been newly introduced by developers during the bug fixes. Combining the two results, we consider that if the developers put much attention to fixing the bugs reported by the tools, then the fix rate of the bugs should be high and the number of newly introduced bugs should be small.\par
When analyzing whether a bug has been fixed or not, we first identify the version file where the bug first appeared. Then, starting from this version file, we analyze the later version files with the results from Mythril and Slither to determine whether the bug has been fixed. Once the bug does not exist in the next versions of the Solidity file, we consider that it has been fixed. For those bugs that exist from the initial version where the bug first appeared to all of the later versions, we consider them to be non-fixed. We compare the time of bug-fix commits to determine whether the bugs are newly introduced or not. If it has not appeared in a previous commit, we assume that it is a newly introduced bug. Figure 12 illustrates the distribution of fixed, non-fixed, and newly introduced bugs, while its horizontal axes show the abbreviation of bugs and the vertical axes show the number of bugs.  \par
By comparing the results in Figure 12, we notice that for most of the 14 categories of bugs, the number of newly introduced bugs exceeds that of the fixed bugs. For some bugs, such as \emph{Uninitialized variables} and \emph{Time manipulation}, the number of newly introduced bugs is close to that of fixed bugs. It shows that the developers may not put much attention to fixing them completely or avoid introducing them again. Because if the developers attach great importance to fixing the bugs reported by the tools completely and avoiding introducing them again, the number of newly introduced bugs should be much smaller than that of the fixed ones. It seems that although there are serious smart contract security incidents \mbox{\citep{SIREN}} \mbox{\citep{BeautyChain}} that are exploited by the bugs like \normalem\emph{Arithmetic} in smart contracts. Maybe these bugs are common in Solidity files, and the developers consider that it is rare and hard to exploit them. So they do not pay much attention to fixing these bugs completely or avoid introducing them strictly. However, \normalem\emph{Access Control} has a different trend from the other categories of bugs. It has the highest fixed percentages and the smallest number of newly introduced bugs, which indicates that developers may pay much attention to fixing them completely and avoid introducing them again when they are fixing this category of bugs. This phenomenon inspires us to think about the priority to fix bugs in the future. There have been many analysis tools proposed to detect bugs in Solidity files. Although they can detect quite a few bugs, developers may not pay much attention to fixing all of them completely. These tools may provide suggestions for developers to fix the bugs with a high fixed percentage and low introduction number like \normalem\emph{Access Control} first.\par

\begin{figure}
	\centering
		\includegraphics[width=0.75\linewidth]{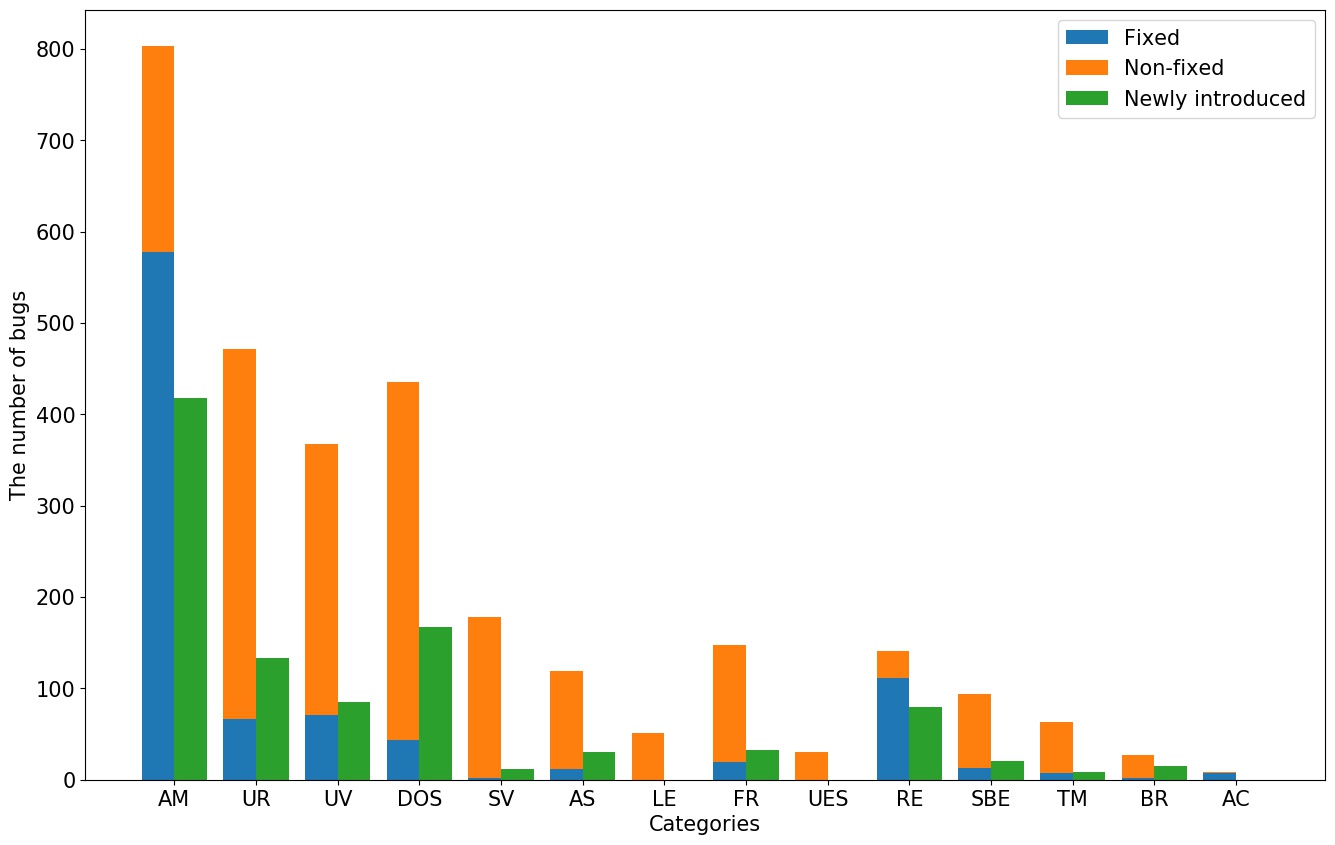}
	    \caption{The number of fixed, non-fixed, and newly introduced bugs during bug fixes. For each category: AM means \protect\normalem{\emph{Arithmetic}}, UR means \protect\normalem{\emph{Unused-return}}, UV means \protect\normalem{\emph{Uninitialized variables}}, DOS means \protect\normalem{\emph{Denial of Service}}, SV means \protect\normalem{\emph{Shadowing variables}}, AS means \protect\normalem{\emph{Arbitrary send}}, LE means \protect\normalem{\emph{Locked-eth}}, FR means \protect\normalem{\emph{Front-Running}}, UES means \protect\normalem{\emph{Unmatched ERC20 Standard}}, SBE means \protect\normalem{\emph{Strict Balance Equality}}, TM means \protect\normalem{\emph{Time manipulation}}, RE means \protect\normalem{\emph{Reentrancy}}, BR means \protect\normalem{\emph{Bad Randomness}}, AC means \protect\normalem{\emph{Access Control}}.}
\end{figure}

\textbf{RQ10.} How do developers fix these bugs?\par
In this part, we try to explore how the developers fix those fixed bugs in RQ9 by manually checking the fix in Solidity files. More specifically, we find the two versions of the Solidity file before and after the bug fix, and then compare their difference to determine the specific fix. Considering the size of the samples, we only select the top 4 bug categories in terms of the total number of fixed bugs. Then, we randomly select a sample of 50 in each of these four categories for manual inspection, and a total of 200 samples. We invite three volunteer graduate students to participate in the manual check process. They have an average of 1.5 years of development experience in Solidity smart contracts. Two of them check the human-written patches independently. If there is a discrepancy between them, the rest participant determines the final result. \par
During our study, we find that these fixed instances can be roughly divided into three categories: one is that the original bug is modified into another one. Developers may change the \normalem{\emph{code}} and \normalem{\emph{function}} of \normalem{\emph{<category, project, function, code>}} to fix the functionality of the smart contracts rather than fix the bug. In this case, we consider that the original bug is modified into a newly introduced bug. The second one is that the bug is deleted. The third is that the bug is truly fixed.\par

\textbf{Reentrancy.} Although we have only considered the \normalem{\emph{Reentrancy}} bugs that are both reported by Mythril and Slither to reduce the false positives. But among our manual check, we find that most of these samples are still false positives. Most of the 50 samples are related to the function \normalem{\emph{addr.transfer()}}, which is marked as a \normalem{\emph{Reentrancy}} bug by Slither. \normalem{\emph{addr.transfer()}} has a built-in gas limit of 2,300 to prevent Reentrancy problems \mbox{\citep{ren2021empirical}}. Some of the published tools mark it as true bugs during their evaluation. But it can not be exploited in practice.\par
\textbf{Arithmetic.} There are 9 false positives in our samples of \normalem{\emph{Arithmetic}}. 18 of the rest samples are modified into the newly introduced bugs. 5 bugs are fixed by deletion. Most of the 18 truly fixed samples are fixed by using a vetted safe math library for arithmetic operations like the example in Listing 4.\par 
\begin{lstlisting}[language=Solidity, frame=none, breaklines=true, literate={\ \ }{{\ }}1,style=Highlight,caption={A fixed example of \protect\normalem{\emph{Arithmetic}}} ]
<   totalSupply += msg.value;
---
>   totalSupply = safeAdd(totalSupply, msg.value);

\end{lstlisting}
\textbf{Uninitial Vairbales.} \normalem{\emph{Uninitial Vairbales}} is caused by using uninitialized storage variables, uninitialized state variables, and uninitialized local variables. These uninitialized variables sometimes may cause financial loss. For example, the developer defines a state variable of address but fails to initialize it. Once the developers use it as the destination to transfer, the Ethers are sent to the address \normalem{\emph{0x0}} and are lost. However, among the 50 samples that we select from \normalem{\emph{Uninitial Vairbales}}, only 18 of them are truly fixes like the example in Listing 5, where the developer set the exact value to initialize the variable \normalem{\emph{liquidationIncentivePercent}}. 23 of them are deleted and the rest 9 are modified into the newly introduced bugs.\par

\begin{lstlisting}[language=Solidity, frame=none, breaklines=true, literate={\ \ }{{\ }}1,style=Highlight,caption={A fixed example of \protect\normalem{\emph{Uninitialized variables}}} ]
<     uint256 public liquidationIncentivePercent;
---
>     uint256 public liquidationIncentivePercent = 5 * 10**18; // 5% collateral discount
\end{lstlisting}

\textbf{Unused-return.} \normalem{\emph{Unused Return}} is caused by the lack of a check on the return value of a function. Among the 50 samples we select, only 6 of them are truly fixed by developers. All of these truly fixed bugs are ensured the return values of the function calls by \normalem{\emph{RequireStatement}} like the example in Listing 6.37 of the rest are modified into the newly introduced bugs. 7 bugs are deleted. \par

\begin{lstlisting}[language=Solidity, frame=none, breaklines=true, literate={\ \ }{{\ }}1,style=Highlight,caption={A fixed example of \protect\normalem\emph{Unused-return}} ]
<   token.transfer(transaction.receiver, _amountReimbursed);
---
>   require(token.transfer(transaction.receiver, _amountReimbursed) != false, "The transfer function must not failed.");

\end{lstlisting}

\textbf{\emph{Finding.4 }} The developers may not put much attention to fixing the bugs reported by the tools completely or avoid introducing them again. Mythril and Slither perform poorly in detecting \normalem{\emph{Reentrancy}} bugs in practice.\par




\subsection{Discussion}
In this section, we provide the actionable implications based on our findings for researchers in Solidity smart contracts.\par

\textbf{Automatic Repair Techniques.} Our findings provide two actionable implications for researchers in the automatic repair techniques of smart contracts. First, it is meaningful to improve the code comment generation method for Solidity files. \normalem{\emph{Comment}} is the most common fixed element. \normalem{\emph{<Comment, addition>}} is the most common fix operation in Solidity files during bug fixes, which indicates the space to research the code comments in Solidity files. The researcher can explore the outdated comments and missing comments in Solidity files. Second, current automatic repair techniques need to extend to multiple-element modifications to fix more bugs. 59\% of the Solidity files involve multi-element modification. Most of the current automatic repair techniques are restricted to one code element, which is not enough to fix the bugs in practice. It is meaningful for automatic repair techniques to learn the relationship between the code elements like \normalem{\emph{EventDefinition}} and \normalem{\emph{EmitStatement}}, and \normalem{\emph{ImportDirective}} and \normalem{\emph{ContractDefinition}} to fix more bugs automatically.\par
 
\textbf{Analysis Tools.} Our results and findings provide three actionable implications of the analysis tools of bugs in Solidity files. First, our results show the complex distribution of bugs at the file level in Solidity smart contract projects. Nearly 40\% of the bug fixes in Solidity files modify two or more Solidity source files. 92\% of the 6,146 bug fixes involve at least one the other source code file. So it is meaningful for researchers to propose analysis tools to check the code link among the Solidity files and to help developers find out the bugs in the other source code files. For example, considering the main usage of the other source code files, researchers can check the function calls from Solidity files used in the other source code files and provide suggestions to modify the other source code files in time while modifying the corresponding code in the Solidity files. It can help developers to determine the impact of bugs more quickly at the different language file levels and reduce the bugs in the other source code files caused by code inconsistencies. Second, although the current analysis tools of bugs in Solidity files can detect quite a few bugs, developers may not pay much attention to fixing all of them completely or avoid introducing them again. So the analysis tools can provide the priorities to fix bugs for developers. For example, the analysis tools can suggest fixing the bugs with a high fixed percentage and low introduction rate like Access Control first. Third, it is important to reduce the false positives in \normalem{\emph{Reentrancy}} reported by analysis tools. Although we have only considered the \normalem{\emph{Reentrancy}} bugs that are both reported by Mythril and Slither to reduce the false positives. But most of the samples we select are still false positives, which are hard to exploit in reality. It seems that these analysis tools need to improve the precision rate to detect \normalem{\emph{Reentrancy}} in reality.\par
\textbf{Solidity developers.} Our results and findings provide the following recommendations to Solidity developers. First, it is better to use the exact version of the Solidity compile. Our results find that some of the bug fixes only involve the \emph{PragmaDirective} during bug fixes. It means the developers only change the versions of Solidity compiles. What is more, most of these fixes change the version of the Solidity compile from a range to an exact one. Therefore, it is recommended that developers accurately specify the version of the compiler at the beginning to reduce the possibility of bugs caused by incorrect versions in the future. Second, to improve the security of smart contracts, use safe library functions and \emph{RequireStatement} for condition checks as much as possible. During our exploration of human-written patches, we find that using safe library functions and \emph{RequireStatement} is the most common way to fix the bugs. So, we suggest developers can learn from these instances and use the same way to avoid some security bugs at first.

\section{Related Work}
Multiple studies that focus on bug fixing and smart contracts have been published in recent years. This section summarizes existing studies in relation to our work.\par
\textbf{Empirical Study on Bug Fixes.}
Yin et al. \citep{yin2011fixes} present a comprehensive characteristic study on incorrect bug fixes from large operating system code bases and a mature commercial OS. Their results show that the bug-fixing process can also introduce errors, which leads to buggy patches that further aggravate the damage. Nguyen et al. \citep{nguyen2013study} present a study of the repetitiveness of code changes in the evolution of 2,841 Java projects. Their results show that the repetitiveness of changes could be very high at small sizes and decreases exponentially as size increases. What's more, repetitiveness is higher and more stable in cross-project settings than in within-project ones. Fixing changes repeat similarly to general changes. Zhong et al. \citep{zhong2015empirical} have conducted an empirical study on thousands of real-world bug fixes from five popular Java projects to analyze the links between the nature of bug fixes and automatic program repair. They summarized two key ingredients of automatic program repair: fault localization and faulty code fix, which provide useful guidance and insights for improving the state-of-the-art of automatic program repair. Campos et al. \citep{campos2017common} explore the underlying patterns in bug fixes mined from software project change histories in Java repositories. They characterized the prevalence of the five most common bug-fix patterns in bug fixes. The results showed that developers often forget to add IF preconditions in the code. Bernardi et al. \citep{bernardi2018relation} try to find out the relation between the bug-inducing and fixing phenomenon and the lack of written communication between committers in open-source projects. They perform an empirical study on four open-source projects and find that increasing the level of communication between fix-inducing committers could reduce the number of fixes induced in a software project. Wen et al. \citep{wen2019exploring} conduct the first systematic empirical study to understand the correlations, in terms of code elements and modifications, between a bug’s inducing and fixing commits. Their results show that leveraging the information of bug-inducing commits can significantly boost the performance of existing automated fault localization and program repair techniques. Wang et al. \citep{wang2020examining} aim to investigate the effects of developers' familiarity with bugs on the efficiency and effectiveness of bug fixing. They conduct an empirical study on 6 well-known Apache Software Foundation projects with more than 9000 confirmed bugs. They find that familiarity with bugs has complex effects on bug fixing: the developers may fix the bugs introduced by themselves more quickly, but they are more likely to introduce future bugs when fixing the current bugs.\par

\textbf{Smart Contract Security.}
In recent years, we have seen a great deal of academic and practical interest in the topic of bugs in smart contracts. To alleviate the problem of insecure smart contracts, researchers have developed various analysis tools and verification frameworks to detect bugs. Oyente \citep{luu2016making} is a symbolic execution tool to find potential security bugs in smart contracts on the Ethereum system. Mythril \citep{Mythril} uses static analysis, taint analysis, and concolic execution to detect a variety of security bugs in smart contracts. ContractFuzzer \citep{jiang2018contractfuzzer} is a fuzzing framework to detect the bugs of Ethereum smart contracts. Securify \citep{tsankov2018securify} converts contract bytecode to Datalog and extracts semantic facts. Then it transforms the bugs into a series of compliance and violation patterns to search for unsafe patterns in smart contracts. SmartCheck \citep{tikhomirov2018smartcheck} checks XML-based intermediate representation of Solidity source code against XPath patterns to detect bugs \citep{ren2021making}. Pied-Piper \citep{Tsinghua} is a static analysis tool that constructs CFG based on bytecode and extracts semantic facts to detect potential backdoors hidden in smart contracts \citep{ren2021making}. SmartEmbed \citep{gao2020deep} is a deep learning-based approach for detecting bugs in smart contracts. \par
To reduce the effort of fixing the bugs of smart contracts, various approaches have been proposed to automatically repair programs. Yu et al. \citep{yu2020smart} present a gas-aware automated smart contract repair approach. The repair algorithm is search-based, and it breaks up the huge search space of candidate patches down into smaller mutually exclusive spaces that can be processed independently. The repair approach considers gas usage of contracts when generating patches for detected bugs. Nguyen et al. \citep{nguyen2021sguard} and Zhang et al. \citep{zhang2020smartshield} propose approaches by analyzing the bytecode to fix potentially vulnerable smart contracts automatically. Nguyen et al. \citep{nguyen2021sguard} propose a tool called SGUARD, which first collects a finite set of symbolic execution traces of the smart contract and then performs static analysis on the collected traces to identify potential bugs. Then it applies a specific fixing pattern for each type of bug in the source code. Zhang et al. \citep{zhang2020smartshield} extract bytecode-level semantic information and utilize them to transform insecure contracts into secure ones. Our study shows the distribution of bugs in real-world Solidity smart contract projects and builds the foundation for the automatic repair of smart contracts.\par

\textbf{Empirical Study on Smart Contract.} 
Chen et al. \citep{chen2020defining} conduct the first empirical study by collecting smart-contract-related posts from Ethereum StackExchange as well as real-world smart contracts to understand and characterize smart contract defects. Pinna et al. \citep{pinna2019massive} perform a comprehensive empirical study of smart contracts deployed on the Ethereum blockchain. Their empirical results show the features of smart contracts and smart contract transactions within the blockchain, the role of the development community, and the source code characteristics. The study contributes to understanding the interaction between Smart Contracts and Blockchain and to the knowledge of the main characteristics of contracts written in Solidity. Wan et al. \citep{wan2021smart} perform a mixture of qualitative and quantitative studies with software practitioners who have experience in smart contract development to understand practitioners’ perceptions and practices on smart contract security. Their results show that smart contract practitioners tend to have a higher awareness of security than practitioners in other software areas. Perez et al. \citep{perez2021smart} focus on finding how many of the vulnerable contracts have been exploited. They survey the vulnerable contracts reported by six recent academic projects and find that, despite the amounts at stake, no more than 2\% of them have been exploited since deployment. Hwang et al. \citep{hwang2020gap} conduct an empirical study on Solidity patches and live contracts to understand the current security status of real-world smart contracts. Their results show that many Solidity developers are unaware of the importance of Solidity patches. Durieux et al. \citep{durieux2020empirical} present an empirical evaluation of 9 state-of-the-art automated analysis tools to obtain an overview of the current state of automated analysis tools for Ethereum smart contracts. Zou et al. \citep{zou2019smart} perform an exploratory study to understand the current state and potential challenges developers are facing in developing smart contracts on blockchains compared to traditional software development. Our study provides a multi-faceted discussion on bug fixes that are extracted from the history of real-world Solidity smart contract projects.

\section{Threats to Validity}
Threat to internal validity is related to the tools we use. There are roughly two categories to analyze the bug in smart contracts: static and dynamic analysis. Static analysis tools catch bugs or vulnerabilities without the need to deploy smart contracts, while dynamic analysis tools work in the opposite way. Static analysis tools have been the main focus of research \citep{perez2021smart}. In this paper, we use the static analysis tool Mythril and Slither to analyze smart contracts. A potential threat to the internal validity is related to the fact that there may be some false positives in our results and they may cause some bias. Our future research agenda is to reduce the influence of these false positives. \par
Threat to external validity is related to the commits that are extracted only from open-source Solidity smart contract projects. Our results may not generalize to commercially developed projects or smart contract projects using different programming languages.\par

\section{Conclusion}
Despite numerous efforts in detecting and repairing the bugs in smart contracts, little is known about bug fixes in Solidity smart contract projects. In this paper, we provide a multi-faceted discussion of bug fixes in real-world Solidity smart contract projects. We conduct an empirical study to explore the File type and amount, Fix complexity, Bug distribution, and Fix patches in the historical bug-fixed versions of the 46 Solidity smart contract projects. We finally distill 4 findings and provide actionable implications from three aspects for researchers to improve the current approaches and propose novel approaches for the bug fixes in Solidity smart contracts.\par
\section*{Acknowledgement}
The work described in this paper is supported by the Key-Area Research and Development Program of Guangdong Province (2020B010164002) and the National Natural Science Foundation of China (61976061, 61902441, 61902105).


\bibliographystyle{cas-model2-names}


\nocite{*}

\bio{}
\endbio


\end{document}